\documentclass{aa}
\usepackage[varg]{txfonts}
\usepackage{amsmath}
\usepackage{graphicx}
\usepackage{natbib}
\usepackage{hyperref}
\usepackage{multirow}
\usepackage{arydshln}
\usepackage[normalem]{ulem}
\usepackage{color}
\bibpunct{(}{)}{;}{a}{}{,} 
\newcommand{\mps}{m\,s$^{-1}$}

\begin{document}

\title{Combined helioseismic inversions for 3D vector flows and sound-speed perturbations} 

\author{David Korda\inst{1}
  \and Michal {\v S}vanda\inst{1,}\inst{2}} 

\offprints{David Korda, \\ \email{korda@sirrah.troja.mff.cuni.cz}}

\institute{Astronomical Institute of Charles University, Faculty of Mathematics and Physics, V~Hole\v{s}ovi\v{c}k\'ach 2, Praha 8, CZ-180~00, Czech Republic
  \and Astronomical Institute of Czech Academy of Sciences, Fri\v{c}ova 298, Ond\v{r}ejov, CZ-25165, Czech Republic} 

\abstract
{Time-distance helioseismology is the method of the study of the propagation of waves through the solar interior via the travel times of those waves. The travel times of wave packets contain information about the conditions in the interior integrated along the propagation path of the wave. The travel times are sensitive to perturbations of a variety of quantities. The usual task is to invert for the vector of plasma flows or the sound-speed perturbations separately. The separate inversions may be polluted by systematic bias, for instance, originating in the leakage of vector flows into the sound-speed perturbations and vice versa (called a cross-talk). Information about the cross-talk is necessary for a proper interpretation of results.}
{We introduce an improved methodology of the time-distance helioseismology which allows us to invert for a full 3D vector of plasma flows and the sound-speed perturbations at once. Using this methodology one can also derive the mean value of the vertical component of plasma flows and the cross-talk between the plasma flows and the sound-speed perturbations.}
{We used the Subtractive Optimally Localised Averaging method with a minimisation of the cross-talk as a tool for inverse modelling. In the forward model, we use Born approximation travel-time sensitivity kernels with the Model S as a background. The methodology was validated using forward-modelled travel times with both mean and difference point-to-annulus averaging geometries applied to a snapshot of fully self-consistent simulation of the convection.}
{We tested the methodology on synthetic data. We demonstrate that we are able to recover flows and sound-speed perturbations in the near-surface layers. We have taken the advantage of the sensitivity of our methodology to entire vertical velocity, and not only to its variations as in other available methodologies. The cross-talk from both the vertical flow component and the sound-speed perturbation has only a negligible effect for inversions for the horizontal flow components. Furthermore, this cross-talk can be minimised if needed. The inversions for the vertical component of the vector flows or for the sound-speed perturbations are affected by the cross-talk from the horizontal components, which needs to be minimised in order to provide valid results. It seems that there is a nearly constant cross-talk between the vertical component of the vector flows and the sound-speed perturbations.}
{}

\maketitle

\section{Introduction}
The internal structure of the Sun and the dynamics of plasmas therein are key factors in understanding the solar activity, which in turn influences the interplanetary space (including effects on human infrastructures). Even though processes in the interior of the Sun may be modelled using numerical codes \citep[e.g.][]{ASH,Rempel_2009}, these models must be constrained by observations to ensure that they describe a physical reality to some extent. 

Helioseismology is the only method that allows us to study subsurface layers of the Sun; these layers, at least to some extent, reflect mostly the properties of the convection zone. Knowledge of the structure and dynamics of plasmas in these regions allows us to put important constraints on theories of the solar dynamo and of the formation and evolution of magnetic fields, including the models of sunspots, and others. The models and inferences from helioseismology do not always agree. For instance, the estimate of convective velocity in the convection zone from helioseismology is two orders of magnitude less than that derived from theoretical studies \citep{2012PNAS..10911928H}. These results, however, were not confirmed by an independent investigation by \cite{2015ApJ...803L..17G} when using a different helioseismic analysis. Also, from theoretical considerations \citep{2012ApJ...757..128M} the results of \cite{2012PNAS..10911928H} do not fit into the current view of convection in the Sun. On the other hand, these discrepancies have not been explained until now. Another surprising result is that there are helioseismic indications that the supergranules may have a different depth structure than predicted by theory \citep{2013SoPh..287...71D,2012ApJ...759L..29S,2014SoPh..289.3421D}. Even though both mentioned results are considered to be controversial, there is an obvious need for both theory and helioseismology to reconcile these existing discrepancies. 

Helioseismology is used to study the solar interior via interpretation  of the dispersion relations of acoustic and surface gravity waves. These waves are generated randomly via the vigorous convection in the convection zone. The waves manifest themselves in many different observables, such as intensity maps of the photosphere or Doppler shifts of photospheric spectral lines. The focus of global helioseismology is to infer the background structure of the solar interior. Local helioseismology focuses on the detection and quantification of small localised perturbations of plasma parameters with respect to the reference solar model.

In this study, we focused on the second approach, using the time--distance method \citep{Duvall_1993} to study the perturbations of travel times of waves caused by inhomogeneities of plasma parameters about the reference model. The time--distance method has been a standard method of solar research for several decades. A comprehensive review of methods of local helioseismology, including the time--distance method and results obtained by this method was presented for example in \cite{2010ARA&A..48..289G} or in \cite{2016LNP...914...25K}. The time--distance helioseismology has been used in many applications: for inversions for the solar rotation \citep{Schou_1997}, meridional circulation \citep{Zhao_2004}, perturbations of density, pressure, temperature, sound speed \citep{Tong_2003,Bruggen_2000}, and plasma flows \citep[e.g.][]{Svanda_2013b}. 

\begin{figure}[h]
    \centering
        \resizebox{6cm}{!}{\includegraphics{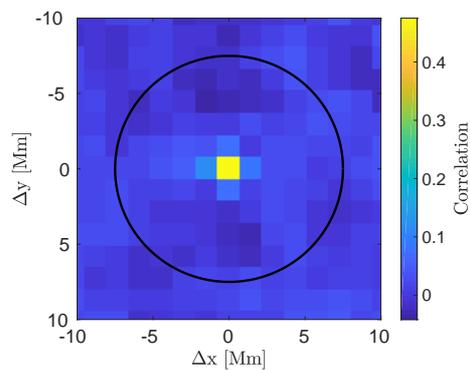}}
        \caption{The map of correlation between the vertical velocity and the sound-speed perturbation in the near-surface layers obtained from the realistic simulation of the solar convection by \citet{Rempel_2014}. This plot is analogous to Fig.~8 of \cite{Svanda_2011}. The black circle corresponds to the width of the target function (15 Mm).}
        \label{pic:dC_vz_corr}
\end{figure}

The usual task of time--distance helioseismology is to invert for the horizontal components and/or spatial variations of the vertical component of the vector of plasma flows, or the sound-speed perturbations separately, using different sets of travel-time measurements. The separate inversions are justified under the assumption of independent action of various perturbers to the wave travel times. In the inverse modelling, however, it is not certain that this assumption holds. Also, in the realistic models of solar plasmas, the perturbers are correlated. For instance, in the mass-conserving convective flows, the upflows (in the vertical direction) in the cell centres are accompanied by outflows (in the horizontal direction). Thus, these two quantities are naturally correlated. This correlation between horizontal and vertical velocities was convincingly shown by \cite{Svanda_2011}; see their Fig.~8. A similar correlation plot between the vertical velocity and the sound-speed perturbation is given in Fig.~\ref{pic:dC_vz_corr}, where a large correlation (correlation coefficient of 0.48 at 0.5 Mm depth) between these two quantities is seen. The correlation coefficient was calculated from a realistic solar convection simulation provided by \citet{Rempel_2014}. This correlation has its origin in the fact that the large upflow velocities are usually located in the interior of the convective cells, where the temperature is greater compared to the surroundings. The greater temperature implies an increase in the sound speed, which naturally follows from the ideal-gas equation of state.

In time--distance helioseismology, the travel times are usually averaged over the annulus around the cell centre, which, together with the correlation described
above, leads to a leakage -- the cross-talk -- of one perturber into the inversion for another one. Separate inversions do not allow one to quantify the possible cross-talk contributions. Information about the cross-talk is extremely important for proper interpretation of results, especially for the vertical velocity \citep{Svanda_2011}. This is because the amplitudes of variations of the vertical velocity are less than 10~\mps{}, but the amplitudes of the horizontal velocities are on the order of 100~\mps{}. Even a small cross-talk can absolutely devalue the results. 

Our aim is to combine inversion for the vector flows $\boldsymbol{v}=(v_x,v_y,v_z)$, where $(x,y,z)$ are local Cartesian coordinates, and for the sound-speed perturbations $\delta c_s$. This approach will allow us to study, for the first time, the cross-talk between the flow perturbations and the sound-speed perturbations. The expected amplitude of the sound-speed perturbations known from the numeric simulations is approximately 50~\mps{}, which is comparable with the amplitude of the vertical velocity, where the cross-talk has a very important effect. 

Another remaining issue in the description of solar flows is that with the current methodology it is possible to invert only for the spatial variations of the vertical velocity. Due to the symmetries of the sensitivity kernels used in the flow inversions, the mean value of the vertical component vanishes. The improved methodology we propose in this paper, that is, merging inversions for the 3D flows and the sound-speed perturbations into one inversion, allows us to fully describe the vertical flow component.

\section{Methodology}
\subsection{Forward problem}
We measure the perturbed travel time $\delta\tau$ by cross-correlating the signal (the Doppler shift of a photospheric spectral line) in a point at a horizontal location $\boldsymbol{r}$ with the signal averaged over the surrounding annulus with the radius $\Delta$ ; we use  the approach of \cite{GB04}. For each $\Delta$, four different geometries are introduced by different weightings in the azimuthal angle (three difference types and one mean type). We use the \textit{outflow-inflow} geometry denoted by o-i, which is the difference between the travel times of the waves travelling from the annulus centre to its rim and the waves travelling in the opposite direction. The \textit{east-west} and \textit{north-south}, denoted by e-w and n-s, are computed similarly, only the signal over the annulus is weighted by either cosine (e-w) or sine (n-s) of the polar angle about the central point. Thus, the measurement is sensitive to the waves travelling in one of these directions. The last one is the \textit{mean} geometry, which averages travel times of the waves travelling from the centre to the rim of the annulus and of the waves travelling in the opposite direction. In the inversions seen in the literature, the difference geometries were used only to invert for flows, whereas the mean geometry was used to invert for the sound speed.  

The perturbed travel time $\delta\tau$ can theoretically be computed from the model as

\begin{equation}
\delta \tau^a \left(\boldsymbol{r}\right)= \int \limits_{\odot} \mathrm{d}^2 \boldsymbol{r}'\, \mathrm{d} z \sum \limits_{\beta = 1}^P K^a_{\beta}\left(\boldsymbol{r}' - \boldsymbol{r}, z\right) \delta q_{\beta}\left(\boldsymbol{r}', z\right) + n^a\left(\boldsymbol{r}\right)\mathrm{,}
\label{eq:dtau}
\end{equation}
where $\boldsymbol{r}$ and $\boldsymbol{r}'$ are the horizontal positions,  $z$ is the vertical position, and $K^a_{\beta}$ are the sensitivity kernels from the forward modelling \citep{GB04,BG07}. Here we use Model S \citep{Model_S} as a reference background model. The $\delta q_{\beta}$ are the perturbations of the physical quantities (such as flows, sound-speed, density, etc.; indexed by $\beta$), $P$ is the number of these quantities, and $n^a$ is a realisation of the random noise \citep{Jackiewicz_2012}. The superscript $a$ corresponds to the measurements of individual travel times and associates the selection of wave filters and averaging geometries as described above, including a selection of the radius of the annulus.

\subsection{Inverse problem}
The goal of an inverse problem is to invert Eq. (\ref{eq:dtau}) and compute

\begin{equation}
\delta q_{\alpha} \left(\boldsymbol{r}_0; z_0\right) = \delta q_{\alpha} \left(\delta \tau^a, K^a_{\beta}\right)
\end{equation}
at a position $\left( \boldsymbol{r}_0; z_0\right)$, where $\alpha$ indicates the perturber (for instance the vertical flow velocity). For simplicity we  use a subscript $\alpha$ to indicate the quantity in the direction of the inversion and a subscript $\beta$ as an index indicating a general physical quantity component. The above illustrated inversion is generally not possible because of high levels of noise. The time--distance helioseismology methods are used to derive an estimate of the given quantity, denoted as $\delta q_{\alpha}^{\mathrm{inv}} \left(\boldsymbol{r}_0; z_0\right)$.

In our methodology we focus on a method called Optimally Localised Averaging (OLA), which corresponds to the method of an approximate inverse in mathematical terminology \citep{louis1990}. The OLA method \citep{Backus_1968} was originally developed for Earth seismology but can also be used for helioseismology \citep{SOLA}. 

\subsubsection*{The SOLA method}
For our study we use an improved version of the Subtractive OLA method (SOLA) described by \citet{Jackiewicz_2012}. The SOLA searches for $\delta q_{\alpha}^{\mathrm{inv}}$ at a given position $\left( \boldsymbol{r}_0; z_0\right)$ in the form of a linear combination of the travel times and the unknown weight functions $w^{\alpha}$:

\begin{equation}
\delta q_{\alpha}^{\mathrm{inv}} \left( \boldsymbol{r}_0; z_0\right) = \sum \limits_{i = 1}^{N} \sum \limits_{a = 1}^{M} w^{\alpha}_{a} \left(\boldsymbol{r}_i - \boldsymbol{r}_0;z_0\right)\delta \tau^a\left(\boldsymbol{r}_i\right)\mathrm{,}
\end{equation}
where $N$ is the total number of the horizontal positions and $M$ is the number of the travel-time geometries. The functions $w^{\alpha}$ minimise the $\chi^2$ in the form
\begin{align}
\chi^2 &= \int \limits_{\odot} \mathrm{d}^2 \boldsymbol{r}'\, \mathrm{d}z \sum\limits_\beta  \left[\mathcal{K}^{\alpha}_{\beta} \left(\boldsymbol{r}', z; z_0\right) - \mathcal{T}^{\alpha}_{\beta}\left(\boldsymbol{r}', z; z_0\right)\right]^2 + \nonumber\\
&+ \mu \sum \limits_{i,\,j,\,a,\,b} w^{\alpha}_a \left(\boldsymbol{r}_i; z_0\right) \Lambda^{ab} \left(\boldsymbol{r}_i - \boldsymbol{r}_j\right)w^{\alpha}_b \left(\boldsymbol{r}_j; z_0\right) + \nonumber\\
&+ \nu \sum \limits_{\beta \neq \alpha} \int \limits_{\odot} \mathrm{d}^2 \boldsymbol{r}'\, \mathrm{d}z \left[\mathcal{K}^{\alpha}_{\beta} \left(\boldsymbol{r}', z; z_0\right)\right]^2 +\epsilon \sum \limits_{a,\,i} \left[w^{\alpha}_a \left(\boldsymbol{r}_i; z_0\right)\right]^2 + \nonumber \\
&+ \sum \limits_{\beta} \lambda^{\beta} \left[\int \limits_{\odot} \mathrm{d}^2 \boldsymbol{r}'\, \mathrm{d}z \mathcal{K}^{\alpha}_{\beta} \left(\boldsymbol{r}', z; z_0\right) - \delta^{\alpha}_{\beta}\right] \mathrm{.}
\label{eq:chi}
\end{align}
We use $\mathcal{T}^{\alpha}_{\beta}=\mathcal{T}\delta_{\beta}^\alpha$, where $\mathcal{T}$ is a user-selected target function in the direction of the inversion and $\delta_\beta^\alpha$ is a Kronecker delta. The target function is localised around the area of investigation. $\Lambda^{ab}\left(\boldsymbol{r}_i - \boldsymbol{r}_j\right) = \mathrm{cov}\left[ n^a\left(\boldsymbol{r}_i\right), n^b\left(\boldsymbol{r}_j\right)\right]$ describes a noise covariance matrix between observations denoted by $a$ and $b$ in positions denoted by $r_i$ and $r_j$. Parameters $\mu$, $\nu$, and $\epsilon$ are user-selected trade-off parameters. The quantity $\mathcal{K}^{\alpha}_{\beta} \left(\boldsymbol{r}', z; z_0\right)$ is defined as

\begin{equation}
\mathcal{K}^{\alpha}_{\beta} \left(\boldsymbol{r}', z; z_0\right) = \sum \limits_i \sum \limits_a w^{\alpha}_a \left(\boldsymbol{r}_i - \boldsymbol{r}_0; z_0\right) K^a_{\beta}\left(\boldsymbol{r}' - \boldsymbol{r}_i, z\right)
,\end{equation}
and is is referred to as the \emph{averaging kernel}. It quantifies the level of smearing of the real quantity $\delta q_{\beta}$. By using the averaging kernels, the real quantity and the estimate from the inversion are linked by
\begin{equation}
\delta q_{\alpha}^{\mathrm{inv}} \left( \boldsymbol{r}_0; z_0\right)=\sum\limits_{\beta} \int \limits_{\odot} \mathrm{d}^2 \boldsymbol{r}'\, \mathrm{d}z\, \mathcal{K}^{\alpha}_{\beta} \left(\boldsymbol{r}' - \boldsymbol{r}_0, z; z_0\right) \delta q_{\beta} \left(\boldsymbol{r}',z\right)+\mathrm{noise}\ .
\label{eq:inv_by_akern}
\end{equation}

The first term of Eq. (\ref{eq:chi}) is the misfit term between the user-selected localisation of the inverted quantity $\delta q_{\alpha}^{\mathrm{inv}}$ and the real localisation given by $\mathcal{K}^{\alpha}_{\beta}$. The second term is the regularisation of the propagation of the random noise. The third term is the regularisation of the cross-talk between the inverted quantity and the other quantities and its meaning is obvious from Eq. (\ref{eq:inv_by_akern}), where the quantity in the direction of the inversion and also other quantities contribute to the inverted estimate. The fourth term minimises the spatial spread of the weight functions and the last term is the normalisation of the averaging kernels added by the Lagrange multipliers $\lambda^{\beta}$ \citep[for details see][]{Svanda_2011}.

The minimisation of Eq. (\ref{eq:chi}) with respect to $w^{\alpha}$ and $\lambda^{\beta}$ gives the solution for the weights. We assume the spatial homogeneity of the background model and therefore the problem is solved in the Fourier domain. There, it decouples and allows us to split a large realistic real-space problem into a series of small Fourier-space problems. This approach permits a realistic inverse problem to be solved even with a desktop computer. One can find the Fourier-space formulation in \citet{Jackiewicz_2012} or \citet{Svanda_2011}.

\paragraph{Noise matrices}
We use data-driven estimates for the noise matrices $\Lambda^{ab}$. They were computed by using a $1/T$ fitting introduced by \cite{GB04}, where $T$ is the travel-time averaging time. We used a large set of travel-time maps measured for 31 days by the \textit{HMI/SDO} instrument \citep{HMI1,HMI2} in the quiet Sun period from August 16, 2010, to September 15, 2010. Following \citet{Lab} we fitted covariances of the travel times $\delta\tau^a$ and $\delta\tau^b$ by using the function $\mathrm{cov}\left[\delta\tau^a,\delta\tau^b\right] = c + d/T$ for $T=\{2, 3, 4, 6, 8, 12, 24\}$~hours, where $c$ and $d$ are the fitted parameters. The constant $d$ is then the element of $\Lambda^{ab}$. We also tried to fit the functions $d/T + e/T^2$ and $c + d/T + e/T^2$ with the fitting parameters $c$, $d$, and $e$ as suggested by \cite{Lab}, but the results were not as expected. For instance, the diagonal terms $\Lambda^{aa}$ (the RMS of the travel times) were overly smeared and did not correspond to the measured RMS values. 

The full covariance matrix is important for getting a reasonable estimate of the variation $\sigma^2_{\alpha}$ of the inverted quantity $\delta q_{\alpha}^{\mathrm{inv}}$, which equals

\begin{equation}
\sigma^2_{\alpha} = \sum \limits_{i,\,j,\,a,\,b} w^{\alpha}_a \left(\boldsymbol{r}_i; z_0\right) \Lambda^{ab} \left(\boldsymbol{r}_i - \boldsymbol{r}_j\right) w^{\alpha}_b (\boldsymbol{r}_j; z_0)\mathrm{.}
\end{equation}

\section{Improvements}
The methodology described above is universal and has been used in various inversions. For instance, it was used to study the vector flows \citep[][and following works]{Svanda_2011} and was found useful for sound-speed inversions \citep{Jackiewicz_2012}. These two types of inversions differ in the selection of the appropriate sensitivity kernels in the inverse problem and hence the corresponding travel-time measurements. For the flows the difference geometries were used (i.e. the o-i, e-w, and n-s geometries), whereas for the sound-speed inversions the mean geometry was used. This means that the subscript $\beta$ in Eq.~(\ref{eq:dtau}) incorporated either the vector of the flow only with $\delta q_{\beta} = \left( v_x,v_y,v_z \right)$ or the fractional sound-speed perturbations only with $\delta q_{\beta} = \left( \delta c_s^2/c_s^2 \right)$. 

This is in agreement with recent findings by \citet{Burston_2015} where the authors show in their Table~1 that the horizontal flow components $v_x$ and $v_y$ are largely sensitive to the difference types of the travel times (o-i, e-w, and n-s), whereas $\delta c_s$ is sensitive to the mean type. The vertical flow component $v_z$ is sensitive to both types however; to invert for the full $v_z$ and not only its spatial variations, the mean-type geometry needs to be included.

Inspired by such findings, we modified the working methodology so that the pipeline can be used to perform inversions for the vector flows and the sound-speed perturbations at once. In effect, the subscript $\beta$ in Eq.~(\ref{eq:dtau}) now enumerates all components of the flow and the sound-speed perturbations at once, $\beta = \left(x, y, z, s\right)$. Thus, the vector $\delta q_{\beta} = \left(v_x, v_y, v_z, \delta c_s\right)$. 

Consequently, the travel-time geometries contain all four averaging geometries, the radii of the averaging annuli, and mode filtering -- all of these choices are covered in the aggregate superscript $a$ in the equations.

This approach allows us to invert for the complete vector of the plasma flows and the sound-speed perturbations in one inversion, which has not been done before. 

We note that  variables were selected after consideration of extensive testing of four possible  constructions of the vector $\delta q_{\beta}$:
\begin{enumerate}
\item The first one was $\left( \boldsymbol{v}/c_s,\delta c_s/c_s\right)$. There is a problem with the interpretation of results, because in order to convert the results into physical units, one has to multiply them by the background sound-speed profile, smeared by the averaging kernel. The averaging kernels for the inverted horizontal velocities scaled by the sound speed had a deeper secondary lobe that could not be removed by varying the trade-off parameters, which made the interpretation of results difficult. Because of the deeper lobe the horizontal velocities were not localised around the target depth. The noise estimates targeted by the inversion were biased.

\item Another option was $\left( \boldsymbol{v},\delta c_s\right)$, which we took as the preferred one. This option has many advantages: The quantities are in physical units. The maps of the quantities, the cross-talk, and the localisation of $\delta q_{\alpha}^{\mathrm{inv}}$ are easy to interpret. The averaging kernels do not have the secondary lobe like in the previous case and are well-localised around the target depth. 

\item[3,4.] The other options were $\left( \boldsymbol{v},\delta c^2_s/c_s\right)$ and $\left( \boldsymbol{v}/c_s,\delta c^2_s/c^2_s\right)$. If we selected one of these options we would have to convert $\delta c^2_s/c_s$ or $\boldsymbol{v}/c_s$ and $\delta c^2_s/c^2_s$ to the physical units similarly to option \#1.
\end{enumerate}

The travel-time kernel codes\footnote{We use the {\sc Kc3} code provided by Aaron Birch.} usually compute the kernels for the fractional squared perturbation of the sound speed, that is, for $\delta c^2_s/c^2_s$. Instead of modifying and extensively testing the kernel code, we rather compute the kernels for linear $\delta c_s$ from the fractional squared perturbations using the following approach. 

We write the squared sound-speed perturbation as 
\begin{equation}
\delta c^2_s = c_s^2 - c_{\mathrm{model}}^2 = 2 c_{\mathrm{model}} \delta c_s + \left(\delta c_s\right)^2,
\end{equation}
where $c_{\mathrm{model}}$ is the background sound-speed value. The second term has a negligible magnitude compared to the first term. We then  compute $\delta c_s$ sensitivity kernels in the form

\begin{align}
K_{\delta c^2_s/c^2_s} \frac{\delta c^2_s}{c^2_s} &= K_{\delta c^2_s/c^2_s} \frac{2 c_{\mathrm{model}} \delta c_s + \left(\delta c_s\right)^2}{c^2_{\mathrm{model}} + \delta c^2_s} = K_{\delta c^2_s/c^2_s} \frac{2 c_{\mathrm{model}} \delta c_s}{c^2_{\mathrm{model}}} = \nonumber \\
&= \frac{2 K_{\delta c^2_s/c^2_s}}{c_{\mathrm{model}}} \delta c_s = K_{\delta c_s}\, \delta c_s,
\end{align}
where we neglected $\delta c^2_s$ against $c^2_{\mathrm{model}}$ (the uncertainty is less than 1 \%). 
Finally, we obtain the relation for the sensitivity kernel for $\delta c_s$:
\begin{equation}
K_{\delta c_s} \equiv \frac{2 K_{\delta c^2_s/c^2_s}}{c_{\mathrm{model}}}\ .
\end{equation}

\section{Validation using synthetic data}
\label{sec:syn}
\subsection{Synthetic travel times}
We validate our new methodology using synthetic data. A snapshot from a box\footnote{Available from \url{http://download.hao.ucar.edu/pub/rempel/sunspot_models/Helioseismology/quiet_sun_98x98x18Mm_64x64x32km/}.} of realistic simulation of the convection zone \citep{Rempel_2014,DeGrave_2014} was used. It naturally contains both the complete vector of the plasma flows and the sound-speed perturbations. 

The synthetic forward-modelled travel times were computed using Eq. (\ref{eq:dtau}). We further add a realistic representation of the random noise to the travel times. Having these two separate components, we were able to study various signal-to-noise-ratio situations.

The realisation of the random noise was computed in the Fourier space using a generator of multivariate normal random numbers using the covariance $\tilde{\Lambda}^{ab}(\boldsymbol{k})$ for each wave vector $\boldsymbol{k}$ separately, where $\tilde{\Lambda}^{ab}$ is the Fourier transform of $\Lambda^{ab}$. We forced the expected Fourier-space symmetries so that in the real space the noise realisation had only a real component as expected. By running a large number (10\,000) of realisations and computing the noise covariance matrix again we verified that the random noise realisations have the same statistical distribution as the corresponding $\Lambda^{ab}$. 

\subsection{Comparison with the model}
We applied our improved pipeline to the synthetic travel times and compared the results to the known inputs. By using the synthetic data, we have a complete understanding of the contributions of the individual components of the inversions, that is, all components of the averaging kernels, the noise, and also the expected \textit{ideal answer}. Thanks to the use of the synthetic data, all these components may be validated separately for each quantity $\delta q_{\alpha}=\{v_x,v_y,v_z, \delta c_s\}$. The inversion components to be validated are
\begin{align}
\delta \hat{q}_{\alpha, \beta}^{\mathrm{inv}} \left(\boldsymbol{r}_0;z_0\right) &= \int \limits_{\odot} \mathrm{d}^2 \boldsymbol{r}'\, \mathrm{d}z\, \mathcal{K}^{\alpha}_{\beta} \left(\boldsymbol{r}' - \boldsymbol{r}_0, z; z_0\right) \delta q_{\beta} \left(\boldsymbol{r}',z\right),\\
\mathrm{noise} \left(\boldsymbol{r}_0;z_0\right) &= \sum \limits_{i,\,a} w^{\alpha}_a \left(\boldsymbol{r}_i - \boldsymbol{r}_0; z_0 \right) n^a\left(\boldsymbol{r}_i\right),\\
\delta q_{\alpha}^{\mathrm{inv}} \left(\boldsymbol{r}_0;z_0\right) &= \sum \limits_{\beta} \delta \hat{q}_{\alpha, \beta}^{\mathrm{inv}} \left(\boldsymbol{r}_0;z_0\right) + \mathrm{noise} \left(\boldsymbol{r}_0;z_0\right) \nonumber\\
&= \sum \limits_{i,\,a} w^{\alpha}_a \left(\boldsymbol{r}_i - \boldsymbol{r}_0; z_0 \right) \delta \tau^a\left(\boldsymbol{r}_i\right),\\
\mathrm{ideal\ answer} \left(\boldsymbol{r}_0;z_0\right) &= \int \limits_{\odot} \mathrm{d}^2 \boldsymbol{r}'\, \mathrm{d}z\, \mathcal{T} \left(\boldsymbol{r}' - \boldsymbol{r}_0, z; z_0\right) \delta q_{\alpha} \left(\boldsymbol{r}',z\right).
\end{align}

The \emph{ideal answer} is what we expect to obtain by applying the inversion. The deviations between the \emph{ideal answer} and our result are caused by the \emph{random noise} and the other-than-expected localisation, described by the 4D averaging kernel.

For this particular study, we used only the $f$-mode ridge-filtered travel times. The reason for this is the strong localisation towards the surface of the Sun, which means that the results, if applied to the real data, can directly be confronted with the results obtained by other methods \citep[similarly to e.g.][]{Svanda_2013a}. The disadvantage of using the $f$ mode is that it is not particularly sensitive to the sound-speed perturbations \citep{Burston_2015}. According to the same paper, the acoustic $p$ modes should have an order-of-magnitude better sensitivity to the sound-speed perturbations. To confirm this we performed equivalent inversions using the $p_1$ mode. These results are discussed in Subsection~\ref{subsect:p1}.

\begin{figure*}
        \sidecaption
        \includegraphics[width=12cm]{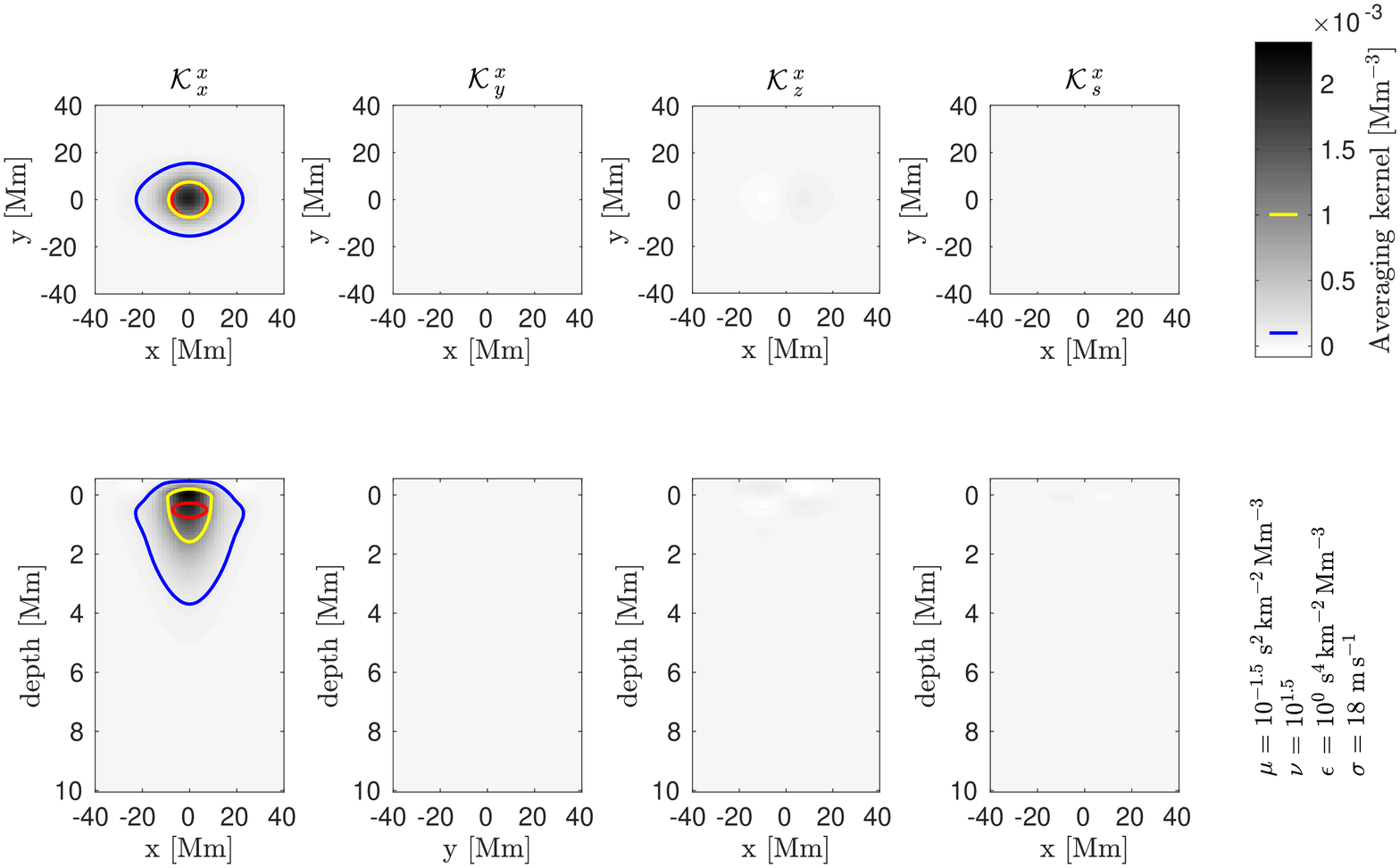}
        \caption{Averaging kernels for $v_x$ inversion at 0.5 Mm depth. The cross-talk was minimised. The red curve corresponds to the half-maximum of the GdG target function, the yellow curve corresponds to the half-maximum of the averaging kernel and the blue solid and blue dotted (not present in this figure) curves correspond to $+5\%$ and $-5\%$ of the maximum of the averaging kernel, respectively. In the top row there are horizontal slices of the averaging kernel at the target depth and in the bottom row there are vertical slices perpendicular to the symmetries. In the right part one can see the selection of the trade-off parameters and the standard deviation when assuming the travel times averaged over 24 hours.}
        \label{pic:vx_akern_GdG}
\end{figure*}
\vspace{1cm}

\begin{figure*}
        \sidecaption
        \includegraphics[width=12cm]{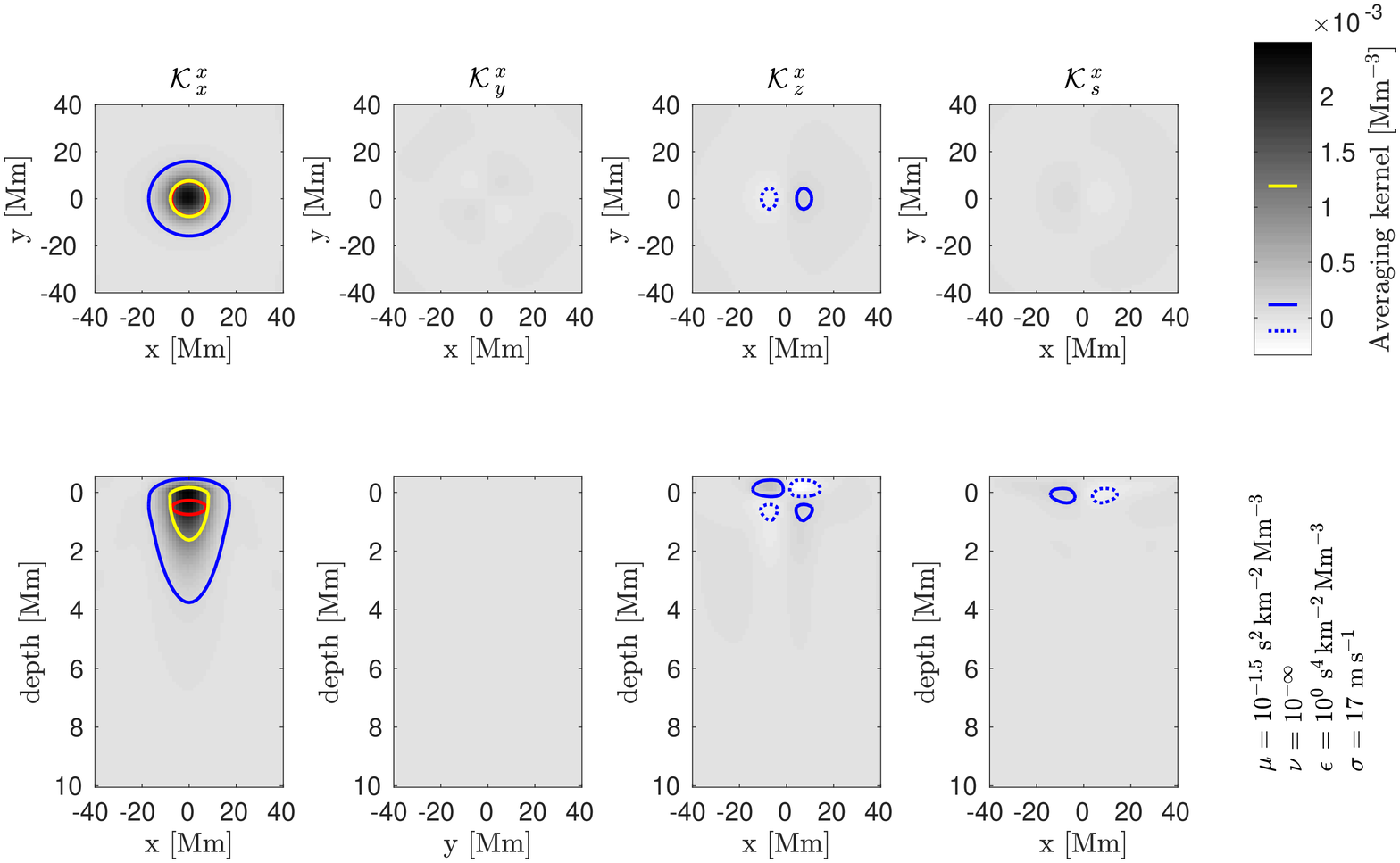}
        \caption{Similar to Fig.~\ref{pic:vx_akern_GdG}: the averaging kernels for $v_x$ inversion at 0.5 Mm depth without cross-talk minimisation.}
        \label{pic:vx_akern_GdG_nu}
\end{figure*}

\begin{figure*}[!h]
        \centering
        \includegraphics[width=17cm]{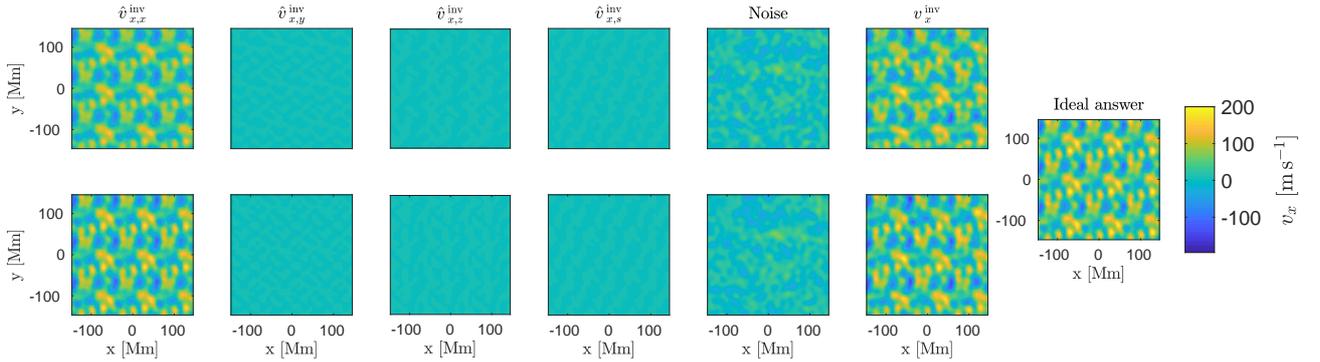}
        \caption{Top row: Inversion with minimisation of the cross-talk. Bottom row: Inversion without minimisation of the cross-talk. First column: $\hat{v}_{x, x}^{\mathrm{inv}}$ contribution to $v_x^{\mathrm{inv}}$. Second column: $\hat{v}_{x,y}^{\mathrm{inv}}$ contribution to $v_x^{\mathrm{inv}}$. Third column: $\hat{v}_{x,z}^{\mathrm{inv}}$ contribution to $v_x^{\mathrm{inv}}$. Fourth column: $\hat{v}_{x,s}^{\mathrm{inv}}$ contribution to $v_x^{\mathrm{inv}}$. Fifth column: the noise contribution to $v_x^{\mathrm{inv}}$. Sixth column: $v_x^{\mathrm{inv}}$. Seventh column: The ideal answer.}
        \label{pic:vx_kostky_GdG}
\end{figure*}

\subsection{The target function} 
In the SOLA inversions, the user is free to select the target function ${\cal T}$, which represents the first-order estimate of the averaging kernel. We describe the results of the comparison with a \emph{Gaussian-damped-Gaussian-type} (GdG) target function, which consists of the 3D Gaussian with the horizontal full width at half maximum (FWHM) of 15~Mm and the vertical FWHM of 0.5~Mm. The centre of the target function is placed at a depth of 0.5~Mm. In the vertical direction, the target function is further multiplied by a smooth function so that ${\cal T}$ reaches zero at the solar surface, $z=0$, and does not extend to the domain of the solar atmosphere. 

\subsection{Trade-off parameters}
The inverted $\delta q_{\alpha}^{\mathrm{inv}}$ depends on a selection of the trade-off parameters $\mu$, $\nu$, and $\epsilon$. We selected the trade-off parameters in the following manner: At first we set an upper limit of the level of the random noise, according to the inverted quantity (20~\mps{} for $v_x$ and $v_y$, 15~\mps{} for $\delta c_s$, 5~\mps{} for $v_z$). This gave us a lower limit of the parameter $\mu$.

In the second step we plotted the averaging kernels and cuts through the weights for all combinations of the trade-off parameters which fulfilled the noise constraint. Our further aims were to minimise the cross-talk which is regularised by the parameter $\nu$. The parameter $\epsilon$ controls the spatial spread of weights to ensure that the inversion weights are spatially confined and to avoid ringing solutions. Larger values of the trade-off parameters are usually balanced by a worse fit of the target function by the averaging kernel. The optimal combination of trade-off parameters is chosen on the trial-and-error basis. We note that there is an \emph{L-curve} \citep[e.g.][]{Hansen_1999} method which attempts to determine the values of the trade-off parameters in an exact way. Unfortunately, in our case this method usually leads to a large level of random noise and therefore we are not using it.

\subsection{The method of evaluation}
In the following sections we discuss the inversions for the horizontal and vertical flow and the sound-speed perturbations separately. When applicable, we compare the new results to the outputs of the \cite{Svanda_2011} pipeline.

For all perturbations we always discuss the shape of the resulting averaging kernel to estimate  the localisation in the Sun, the smearing, and the cross-talk contributions. In the plots we show the horizontal slices through  the target depth and also the vertical slices perpendicular to the symmetry axis. We then describe the individual contributions to the inverted quantity and discuss the fulfilment of the inversion requirements, such as the spatial localisation of the weights.

\section{Inversions for horizontal components of vector flows}
Inversions for horizontal (that is $v_x$ and $v_y$) components of the vector flows have been typical tasks for the time-distance helioseismology. There is only a small amount of cross-talk with the vertical $v_z$ caused by natural correlations between the vertical upflow and the outflows in the mass-conserving flow. The averaging kernels are plotted in Fig.~\ref{pic:vx_akern_GdG} (cross-talk minimised) and Fig.~\ref{pic:vx_akern_GdG_nu} (cross-talk not minimised). (We note that only plots for $v_x$ are given, plots for $v_y$ are similar except for the rotation by 90 degrees around the vertical axis.) By comparing the second, third, and fourth columns of the two figures, one can see that the cross-talk minimisation works as expected, when these contributions vanish in  Fig.~\ref{pic:vx_akern_GdG}. The cross-talk minimisation is balanced by the worse fit of the component of the averaging kernel in the direction of the inversion to the target function. The averaging kernel ${\cal K}_{x}^{x}$ is more extended in the $x$ direction, reaching essentially an elliptical shape in the horizontal cut, whereas the target function is roundish. The use of the $f$-mode sensitivity kernels only also contributed to this issue because they do not represent the GdG function well. 

\begin{figure*}
        \sidecaption
        \includegraphics[width=12cm]{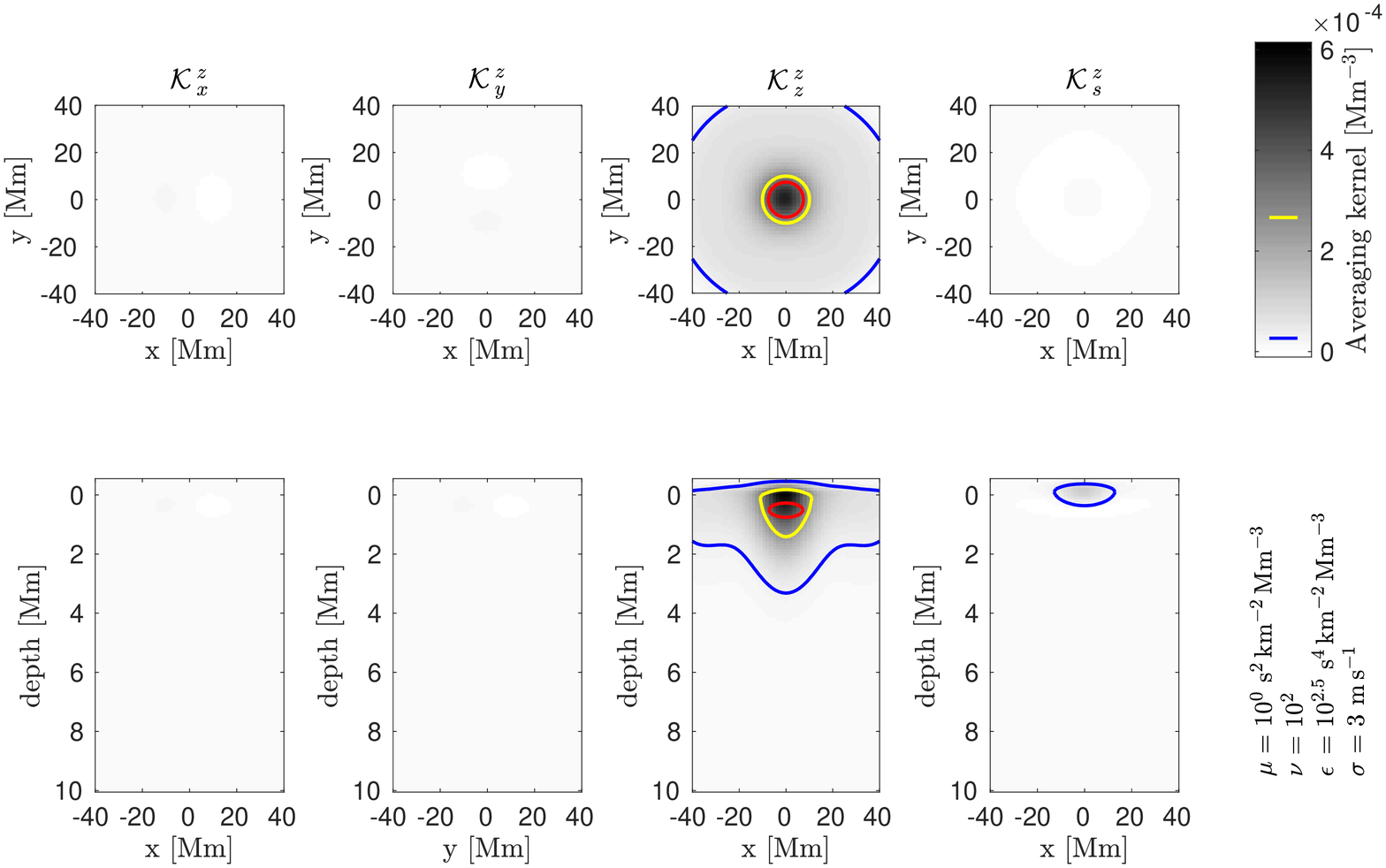}
        \caption{The averaging kernels for $v_z$ inversion at 0.5 Mm depth. The cross-talk was minimised. See   Fig. \ref{pic:vx_akern_GdG} for details.}
        \label{pic:vz_akern_GdG}
\end{figure*}

\begin{figure*}
        \sidecaption
        \includegraphics[width=12cm]{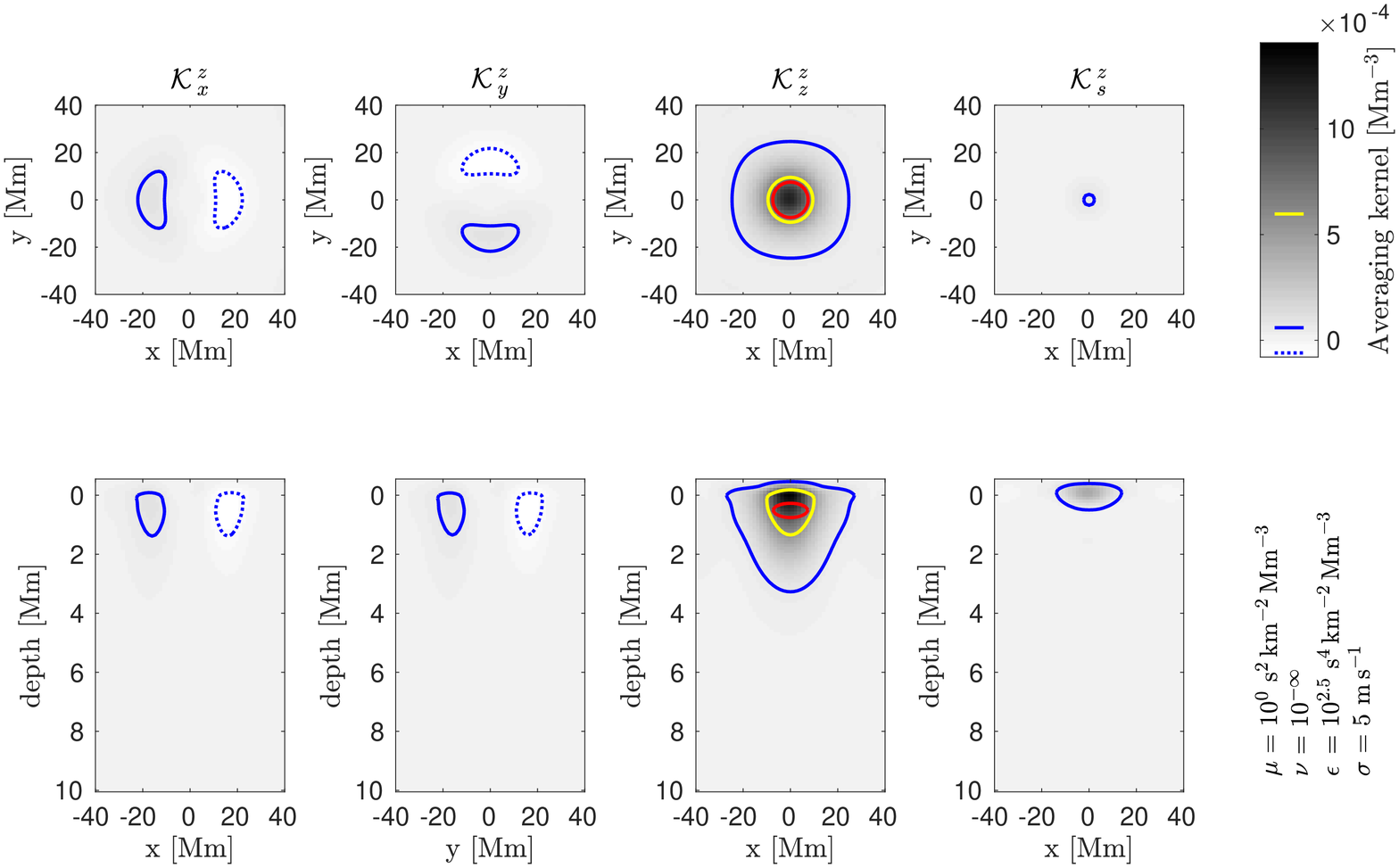}
        \caption{The averaging kernels for $v_z$ inversion at 0.5 Mm depth. No cross-talk minimisation. See   Fig. \ref{pic:vx_akern_GdG} for details.}
        \label{pic:vz_akern_GdG_nu}
\end{figure*}

In Fig. \ref{pic:vx_kostky_GdG} one can see the individual contributions from the different quantities and the noise to both inversions (with and without the minimisation of the cross-talk). There is also a comparison with the ideal answer. Despite the appearance of the averaging kernel with cross-talk not being minimised, the total leakage from $v_z$ and $\delta c_s$ is negligible because the cross-talk components of the averaging kernel have a rather small extent. Moreover, the amplitudes of the leaking components are at least an order of magnitude smaller in the near-surface layers.

Due to the fact that the amplitudes of the horizontal flows are at least an order of magnitude larger than the amplitudes of the two remaining perturbers, there is no need to use the improved methodology for such inversions. However, it might be unnecessary to minimise the cross-talk at all. Such a claim will, however, lose its validity in the layers or regions, where the horizontal and vertical components reach a comparable magnitude.

\section{Inversion for the vertical flow component}
\label{sect:vz}
Until now, only variations of the vertical component $v_z$ were known. This is because only the difference travel-time geometries were used to invert for $v_z$ and the total horizontal integral of the sensitivity kernels vanished due to their symmetries \citep{Burston_2015}. Therefore, such measurements are not sensitive to the horizontal average (the mean) $\langle v_z\rangle$ which is therefore identically zero in the inverted estimates. We only note that this property also requires a modification of the target function for such an inversion \citep{Svanda_2011} in order to fulfil the normalisation constraint in Eq.~(\ref{eq:chi}).

According to \cite{Burston_2015} the mean travel-time geometry is sensitive to $\langle v_z\rangle$. Therefore, by adding the mean geometry we can invert for full $v_z$. From the previous works \citep[e.g.][]{Svanda_2011} it is known that there is a large cross-talk with the horizontal flow components, which must be minimised in order to have credible results.

\begin{figure}[h]
       \centering
        \resizebox{6cm}{!}{\includegraphics{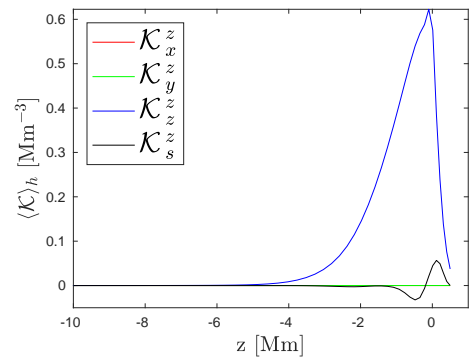}}
        \caption{The horizontal integral of the mean geometry contribution of various physics quantities to the averaging kernel.}
        \label{pic:vz_akern_h_GdG}
\end{figure}

The inversion averaging kernels are plotted in Figs.~\ref{pic:vz_akern_GdG} (cross-talk minimised) and \ref{pic:vz_akern_GdG_nu} (cross-talk unconstrained). The cross-talk minimisation approach allows us to minimise the leakage from the horizontal flow components, however, it has only a limited effect on the leakage of the sound-speed perturbations. We found that this cross-talk with $\delta c_s$ has the origin in the contribution of the mean travel-time kernels. To estimate the severity of this cross-talk, we plotted the mean-geometry contribution of the four physical quantities to the averaging kernel. As one can see from Fig.~\ref{pic:vz_akern_h_GdG}, the mean contribution comes from the ${\cal K}^z_z$ kernel, with a small contribution from the ${\cal K}^z_s$ kernel. There is no contribution from the horizontal flow, because the mean travel-time sensitivity kernels for the horizontal components have vanishing horizontal integrals. From this plot we estimate that the leakage from the sound-speed perturbation to the inversion for the vertical flow is very small.

\begin{figure*}
        \centering
        \includegraphics[width=17cm]{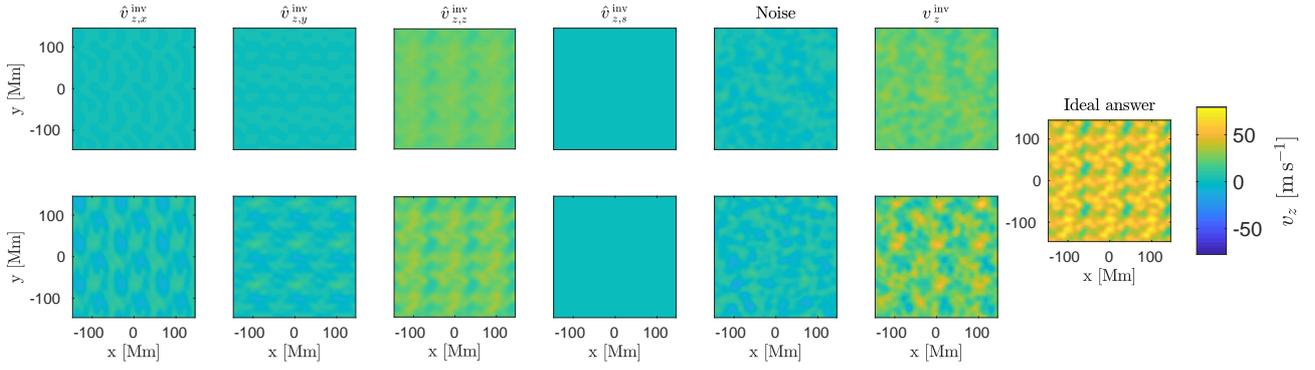}
        \caption{The individual contributions into $v_{z}^{\mathrm{inv}}$ and its comparison with the ideal answer. See Fig. \ref{pic:vx_kostky_GdG} for details.}

        \label{pic:vz_kostky_GdG}
\end{figure*}

\begin{figure}
        \includegraphics[width=0.45\textwidth]{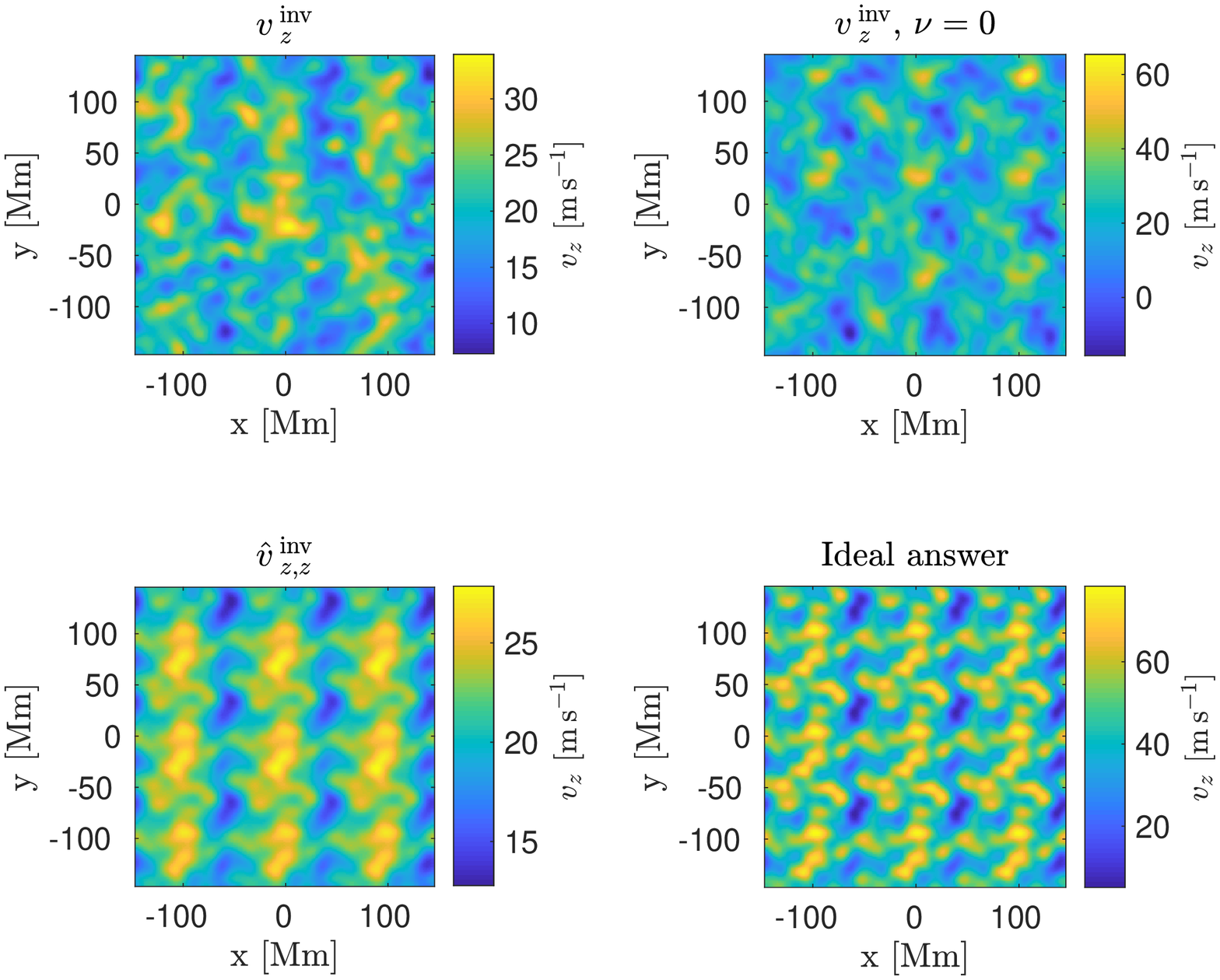}
        \caption{Top row: $v_{z}^{\mathrm{inv}}$ with minimisation of the cross-talk and $v_{z}^{\mathrm{inv}}$ without minimisation of the cross-talk. Bottom row: $\hat{v}_{z,z}^{\mathrm{inv}}$ and the ideal answer. The correlation coefficient between $\hat{v}_{z,z}^{\mathrm{inv}}$ and the ideal answer is 0.87. The colour bars are not the same.}
        \label{pic:vz_comp_GdG}
\end{figure}

The effect of the minimisation of the cross-talk with the horizontal component is clearly demonstrated in Fig.~\ref{pic:vz_kostky_GdG}, where individual inversion components are plotted. The cross-talk contributions seen in the bottom row have a significant impact on the sum and their value in the case of the horizontal-components leakage is higher than the typical variations of $\hat{v}_{z,z}^{\mathrm{inv}}$. For a better comparison between the components and the ideal answer, in Fig. \ref{pic:vz_comp_GdG} we plot $v_{z}^{\mathrm{inv}}$ with and without the minimisation of the cross-talk, $\hat{v}_{z,z}^{\mathrm{inv}}$, and the ideal answer. These plots show that the cross-talk minimisation indeed improves the inversion results. The correlation coefficients between the inverted value and the ideal answer is 0.36 when the cross-talk is ignored and 0.57 when it is minimised. By comparing our new results to the results of \cite{Svanda_2011} we can say that even the supplement of the mean travel times in the vertical flow inversion helped to lower the cross-talk, because without them, the cross-talk unconstrained inversions ended with a correlation coefficient of close to $-1$ with the expected answer \citep[also][]{2007ApJ...659..848Z}. 

Obviously, the inversion scales the values by a factor of 0.44, which is due to a large misfit between the target function and the averaging kernel. Such an issue may be understood as being due to the fact that the inferred vertical flows are smoother than those expected from the setup. Such a result highlights the importance of having the information about the averaging kernel when interpreting the inversions. 

\begin{figure}
        \includegraphics[width=0.45\textwidth]{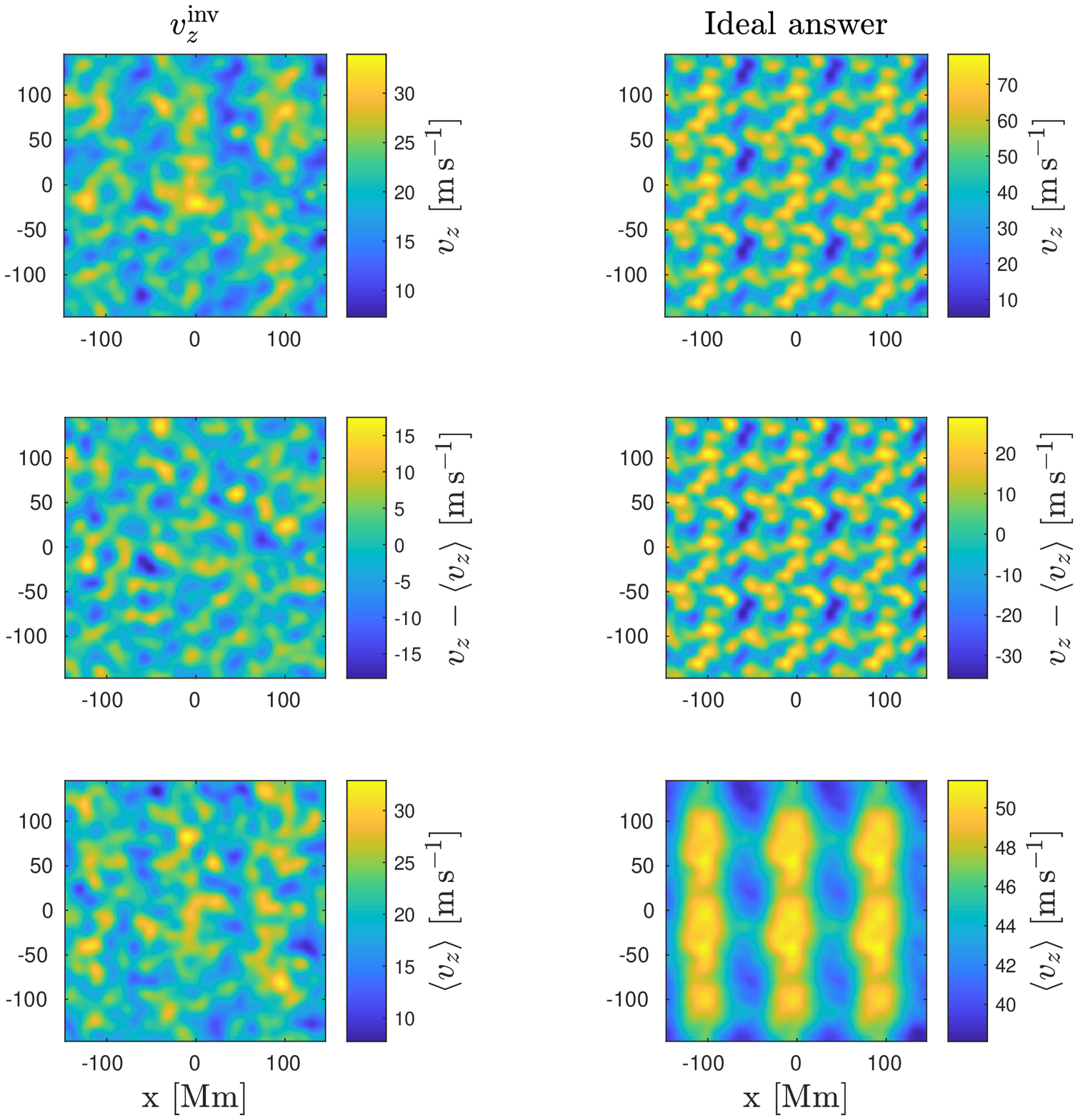}
        \caption{Top row: Results of the inversion for $v_z$ with the mean and the difference geometries. Middle row: Results of the inversion with the difference geometries only. Bottom row: Difference of two aforementioned inversions. Left column: $v_{z}^{\mathrm{inv}}$. Right column: The ideal answers of the inversions. We point out that the colour bars are not scaled.} 
        \label{pic:vz_mean}
\end{figure}

The incorporation of the mean travel times allowed us to invert for the full vertical velocity and not only its horizontal variations. To demonstrate this, we additionally inverted for the vertical velocity using the old pipeline of \cite{Svanda_2011} using only the difference travel-time kernels. It is necessary to modify the target function so that it has a vanishing horizontal mean, which is achieved by introducing a wide negative horizontal side lobe. The results shown in Fig. \ref{pic:vz_mean} demonstrate that the expected ideal answers obtained by using two different target functions are essentially indistinguishable, except for the systematic offset. By subtracting the two maps (seen in the third row of Fig. \ref{pic:vz_mean}) we see that this offset actually has a structure with a very small variation compared to the actual variations of the ideal answer. The offset represents a large-scale structure of the vertical flow, which is not detectable using the \cite{Svanda_2011} approach. 

When the mean travel times are incorporated into the inversion, the overall structure of the inferred vertical flow does not change. The correlation coefficient of the improved vertical flow inversion and the inversion by the original pipeline is 0.91. However the large-scale offset is successfully measured by the improved methodology as one can see from the differences in the bottom row of Fig. \ref{pic:vz_mean}. Except for the lowered amplitude (due to the imperfect averaging kernel), its structure is measured reasonably well.
\begin{figure*}
        \sidecaption
        \includegraphics[width=12cm]{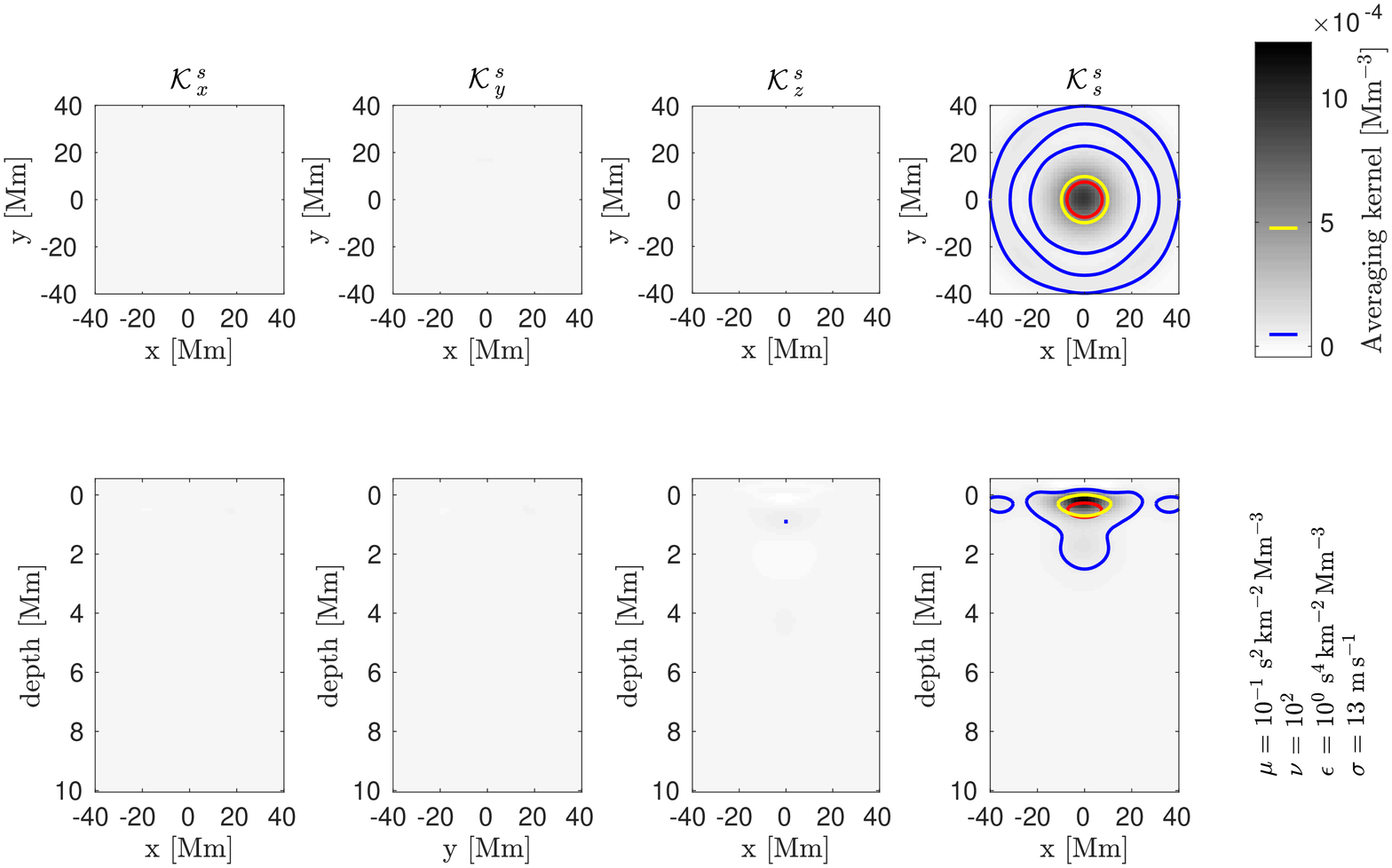}
        \caption{The averaging kernels for $\delta c_s$ inversion at 0.5 Mm depth. The cross-talk was minimised. See Fig. \ref{pic:vx_akern_GdG} for details.}
        \label{pic:cs_akern_GdG}
\end{figure*}

\begin{figure*}
        \sidecaption
        \includegraphics[width=12cm]{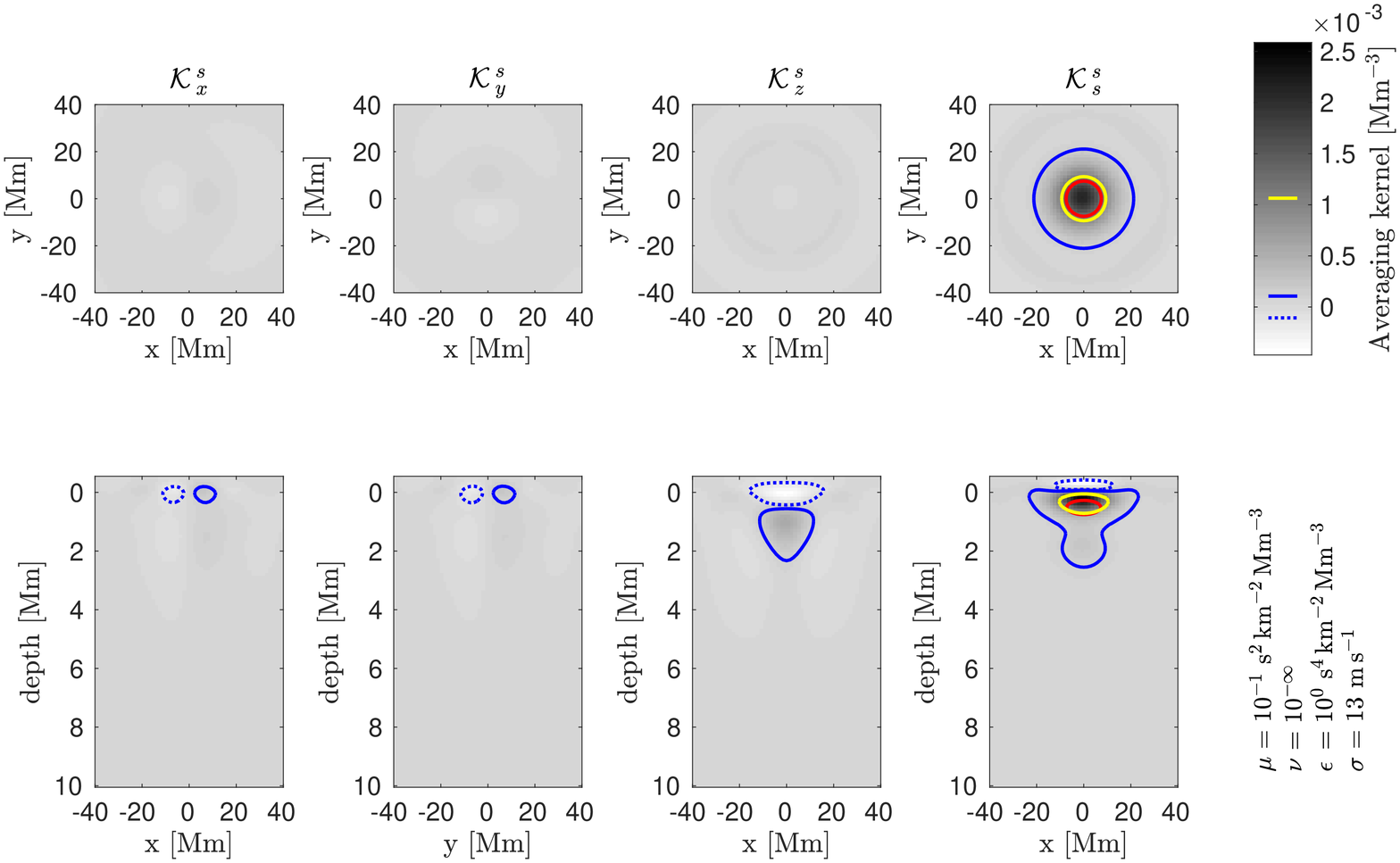}
        \caption{The averaging kernels for $\delta c_s$ inversion at 0.5 Mm depth. No cross-talk minimisation. See Fig. \ref{pic:vx_akern_GdG} for details.}
        \label{pic:cs_akern_GdG_nu}
\end{figure*}

\begin{figure*}
        \centering
        \includegraphics[width=17cm]{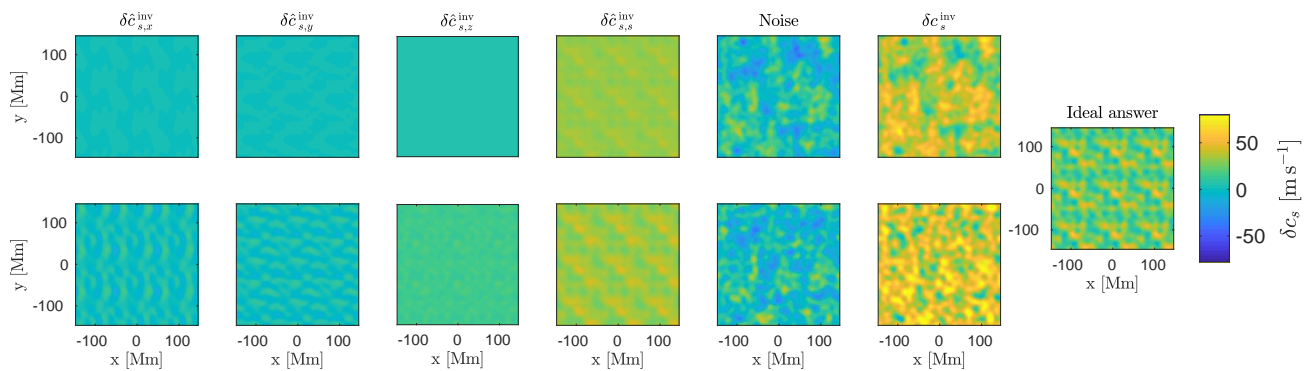}
        \caption{The individual contributions from $\delta \hat{c}_{s,\beta}^{\mathrm{inv}}$ and noise to $\delta c_{s}^{\mathrm{inv}}$ and its comparison with the ideal answer. See Fig. \ref{pic:vx_kostky_GdG} for details.}
        \label{pic:cs_kostky_GdG}
\end{figure*}

\begin{figure}
        \includegraphics[width=\hsize]{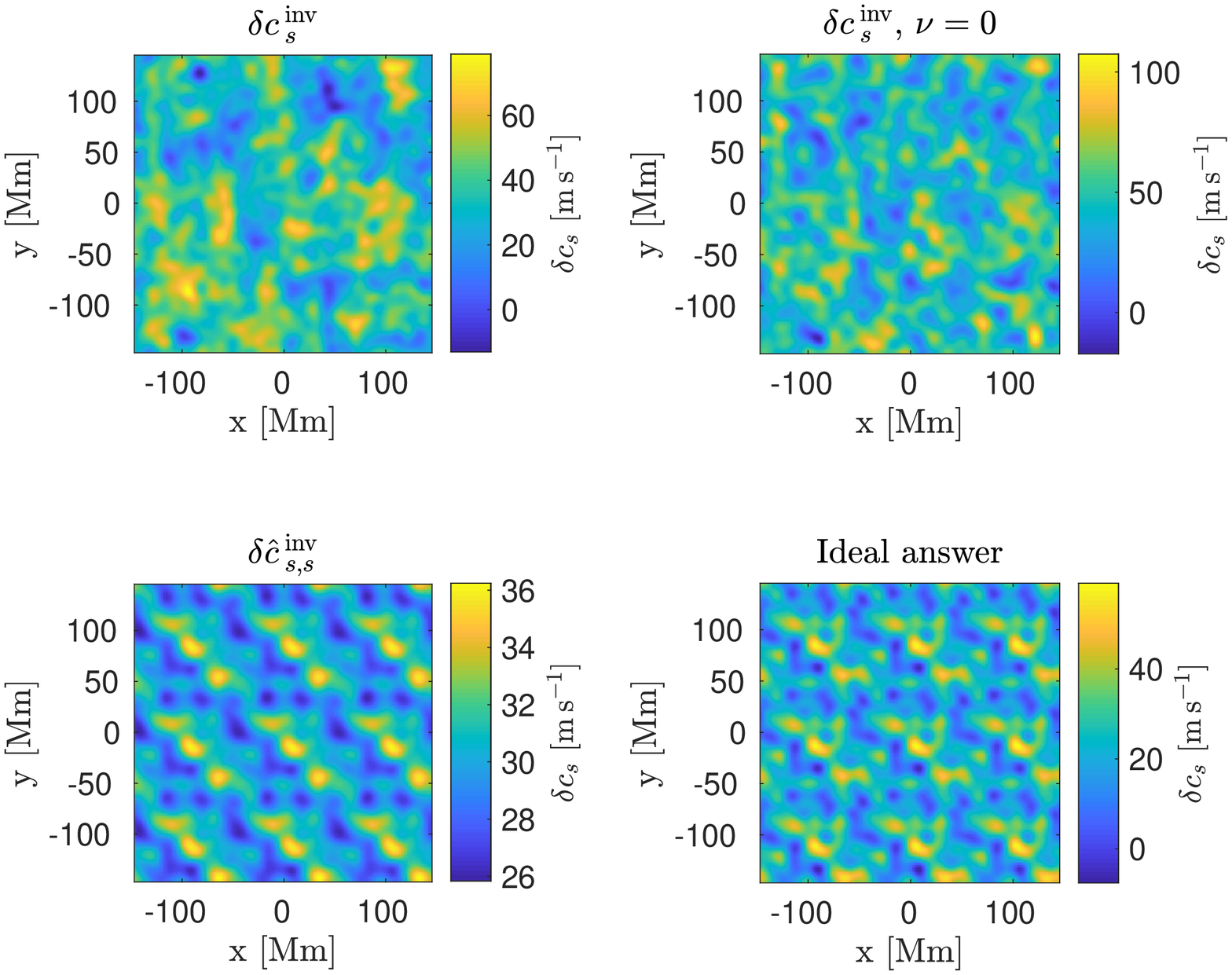}
        \caption{Top row: $\delta c_{s}^{\mathrm{inv}}$ with minimisation of the cross-talk and $\delta c_{s}^{\mathrm{inv}}$ without minimisation of the cross-talk. Bottom row: $\delta \hat{c}_{s,s}^{\mathrm{inv}}$ and ideal answer. The correlation coefficient between $\delta \hat{c}_{s,s}^{\mathrm{inv}}$ and ideal answer is 0.90. The colour bars are not the same.}
        \label{pic:cs_comp_GdG}
\end{figure}

\section{Inversion for sound-speed perturbations}
\label{sect:cs}
An example of the inversion for the sound-speed perturbations using principally the same methodology we use in this study was published by \citet{Jackiewicz_2012}. Their inversion used only the mean travel-time sensitivity kernels and thus did not take into account possible leakage from other quantities. We have the opportunity to do this now.

The sound-speed inversion averaging kernels are plotted in Figs.~\ref{pic:cs_akern_GdG} (with the cross-talk minimisation) and  \ref{pic:cs_akern_GdG_nu} (without the cross-talk minimisation). The fits in the direction of the inversion (the right-most panels) are not very representative of the imposed GdG target function, which is the consequence of using the $f$ mode only. On the other hand, the cross-talk minimisation performs well by essentially setting the cross-talk components of the averaging kernels to zero. For that one has to expect a worse fit of the target function in the direction of the inversion, where even an extended small-amplitude surrounding annulus appears. We assume that the fit improves when more independent travel-time measurements are used in the inversion because of a larger number of independent observations. A wider averaging kernel causes a larger-than-expected smoothing of the results, which is demonstrated by the smaller amplitude of the inverted sound-speed perturbations compared to the ideal answer. 

In Fig. \ref{pic:cs_kostky_GdG} we plot individual contributions to $\delta c_{s}^{\mathrm{inv}}$. In both cases the inverted sound-speed-perturbations maps are dominated by the component in the direction of the inversion, which has the RMS value of 31~\mps. The RMS of the total cross-talk is 16~\mps{} when not minimised. The cross-talk reaches half of the magnitude of the desired signal, which we consider too large. After the minimisation, the cross-talk contribution lowers to only less than 4~\mps. The cross-talk is positively correlated with the inverted value, that is, without its minimisation, we overestimate the real sound-speed perturbations by a factor of about 1.9. 

In Fig. \ref{pic:cs_comp_GdG} one can compare $\delta c_{s}^{\mathrm{inv}}$ with and without the minimisation of the cross-talk, $\delta \hat{c}_{s,s}^{\mathrm{inv}}$, and the ideal answer. The correlation coefficient between $\delta \hat{c}_{s,s}^{\mathrm{inv}}$ and the ideal answer is 0.90. The difference between $\delta c_{s}^{\mathrm{inv}}$ (top left) and the ideal answer is mainly caused by the misfit because the sensitivity of the $f$ mode to $\delta c_s$ is weak.

\subsection*{Validation with the $p_1$ mode}
\label{subsect:p1}
According to \cite{Burston_2015}, the sensitivity of the acoustic modes is at least one order of magnitude higher than that of the $f$ mode. Therefore we ran the sound-speed inversion again, using only the kernels for the $p_1$ ridge. The first acoustic ridge samples similar depths to the $f$ mode, and therefore the inversions are directly comparable. 

Furthermore, in the inversions presented above, we imposed a certain shape (the ``GdG'') of the target function. We showed that in some cases the inverted results were far from optimal, which was due to the fact that the $f$-mode travel-time sensitivity kernels were not dependent enough on depth to fit the GdG target function. For this reason we recomputed the inversion with a target function that has a more natural profile in the vertical domain.

In our methodology, such a target function (called Arbitrary-z, \emph{Az}) resembles the natural depth profile of the sensitivity kernels used in the inversion by representing their kinetic-energy profile. It is usually obtained by simply averaging the horizontal integrals as a function of depth of all sensitivity kernels used in the given inversion. Using the Az-class target function our inversion is almost equivalent to the 2D inversion and thus it is expected that the resulting averaging kernels will better fit this target function, while having a lower level of random noise at the same time. In the horizontal direction the target function is again a Gaussian with a FWHM of 15~Mm. 

\begin{figure*}[h]
        \sidecaption
        \includegraphics[width=12cm]{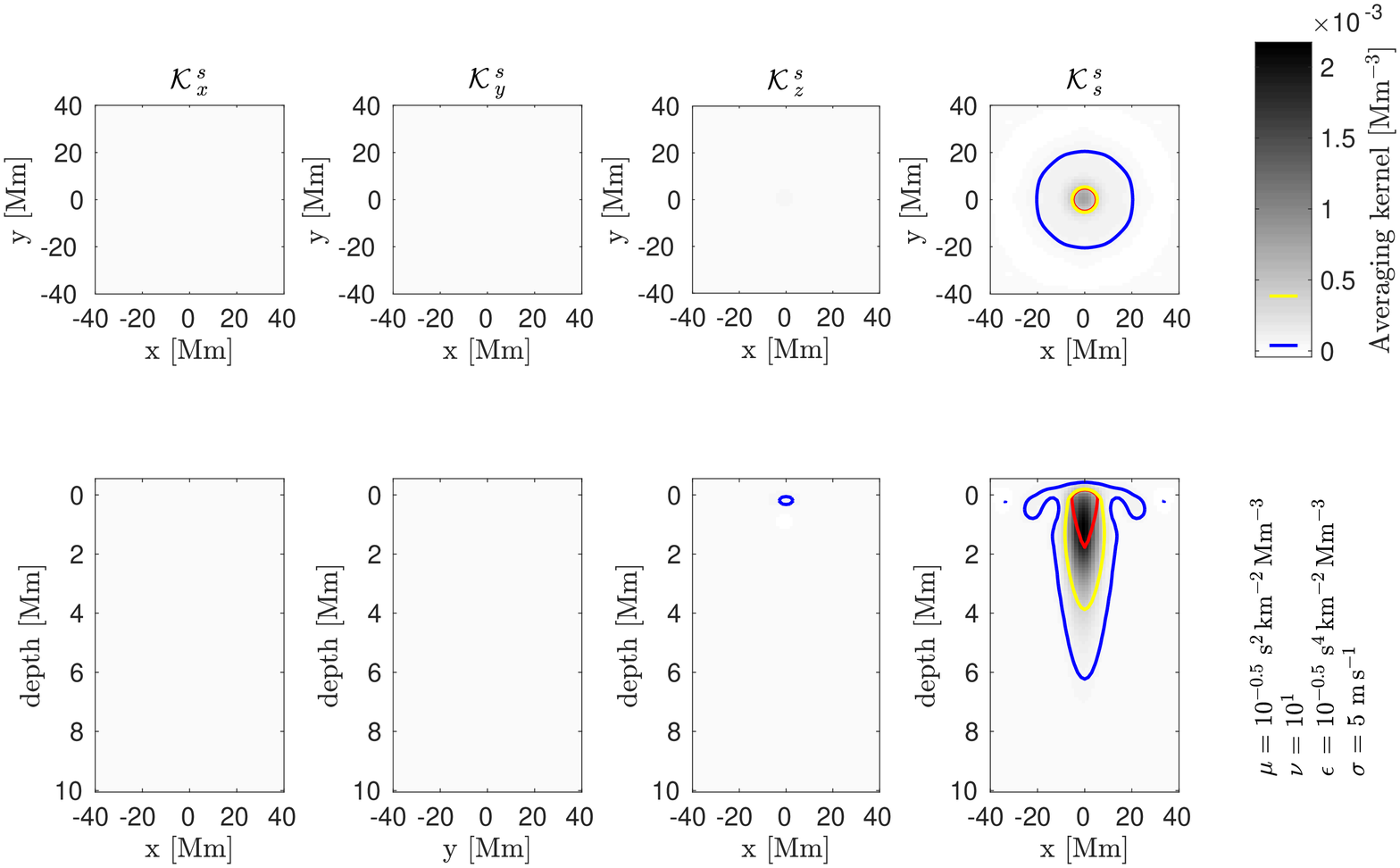}
        \caption{The averaging kernels for $\delta c_s$ inversion at 0.5 Mm depth. No cross-talk minimisation. The Az target function was used. See Fig. \ref{pic:vx_akern_GdG} for details.}
        \label{pic:cs_akern_Az_p1}
\end{figure*}

\begin{figure*}
        \sidecaption
        \includegraphics[width=12cm]{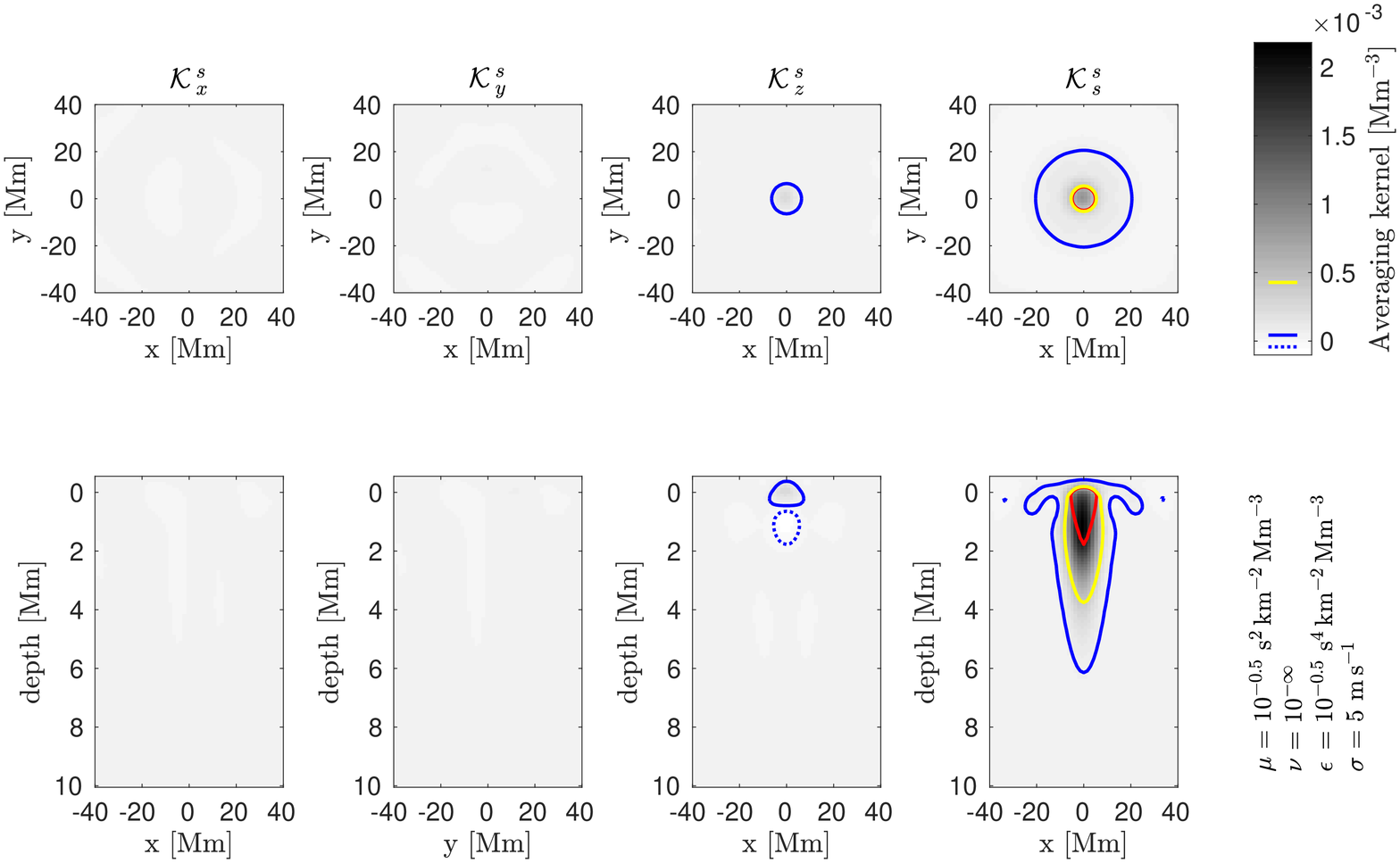}
        \caption{The averaging kernels for $\delta c_s$ inversion at 0.5 Mm depth. No cross-talk minimisation. The Az target function was used. See Fig. \ref{pic:vx_akern_GdG} for details.}
        \label{pic:cs_akern_Az_nu_p1}
\end{figure*}

\begin{figure}
        \resizebox{\hsize}{!}{\includegraphics{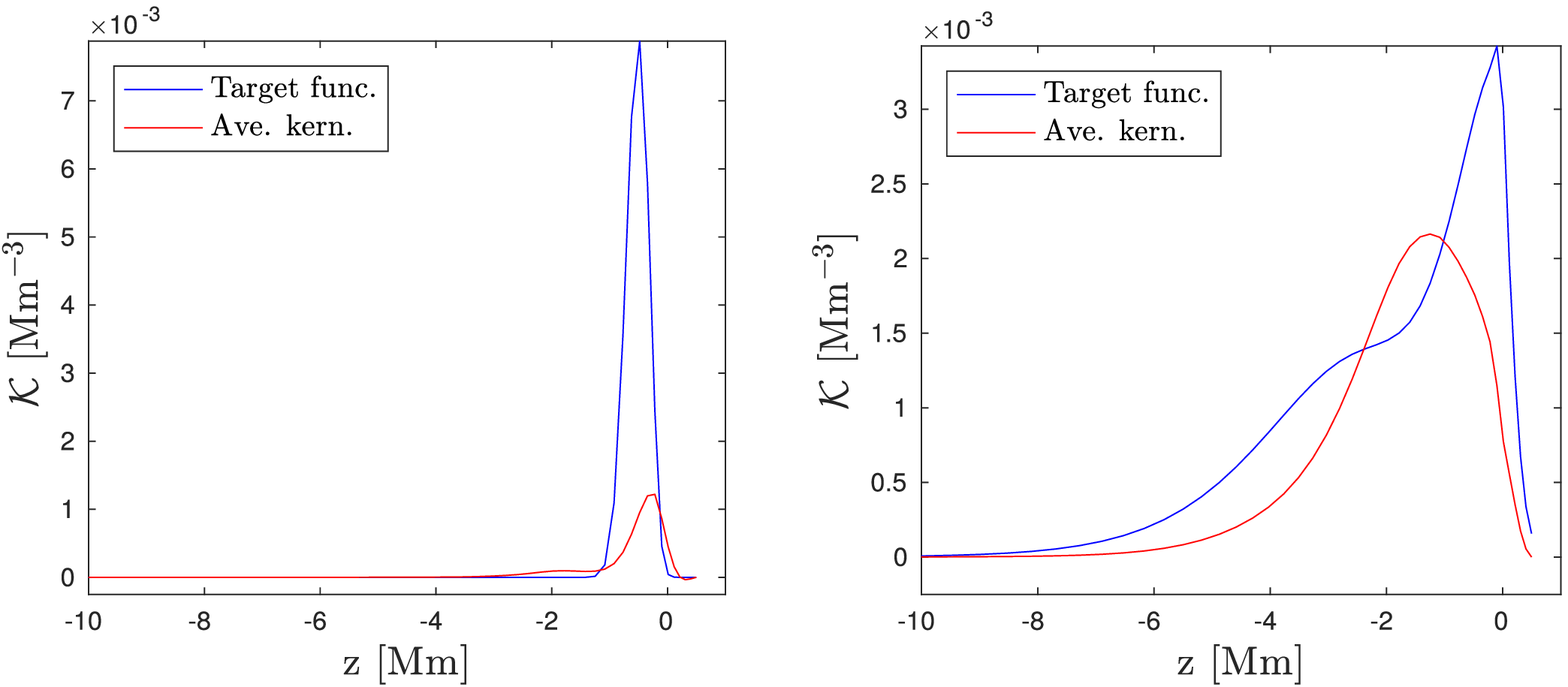}}
        \caption{Left: The cut through the line $x = 0$, $y = 0$ of the ${\cal K}^s_s$ component of the averaging kernel and its comparison with the target function for the GdG $f$-mode inversion. Right: The same as in the left panel but for the Az $p_1$-mode inversion.}
        \label{pic:ra_rt_c_s_Az_p1}
\end{figure}

Figure~\ref{pic:cs_akern_Az_p1} (to be compared with Fig.~\ref{pic:cs_akern_GdG}) demonstrates that the representation of the target function by the averaging kernel is indeed much better in the case of inversion using the Az-type target function and the $p_1$ mode. A similar conclusion may be drawn when comparing Figs.~\ref{pic:cs_akern_Az_nu_p1} and \ref{pic:cs_akern_GdG_nu} for the inversions with an ignored cross-talk. The superiority of the $p_1$ inversion for $\delta c_s$ is even more visible when comparing the horizontally averaged averaging kernels plotted as a function of depth (Fig.~\ref{pic:ra_rt_c_s_Az_p1}). 

\section{Summary}

The results shown in the figures presented here can also be represented by numbers. In Table \ref{tab:GdG} we give some statistical properties of the inversions for all physical quantities in question, that is, for three components of the flow velocity and the sound-speed perturbations. 

For each quantity separately (by columns) we give the correlation coefficient of the inversion component in the direction of the inversion with the ideal (expected) answer, which is always positive and large. The correlation coefficient between the inverted estimate and the ideal answer is given, which is always positive, however, it is lower for the vertical flow and sound-speed perturbations. That is mainly due to the presence of the random noise. Finally, the RMS values of the component in the direction of the inversion, of the total cross-talk contribution, and of the random noise help to evaluate the signal-to-noise ratio of the inverted estimates and the significance of the cross-talk pollution. The comparison between the upper and lower tables demonstrates the expected performance of the cross-talk minimisation.
\begin{table}
\caption{Correlation coefficients and RMS of the inversions using the synthetic travel times. In the first two rows and the eighth and ninth rows there are the correlation coefficients of $\delta \hat{q}_{\alpha, \alpha}^{\mathrm{inv}}$ and the whole inversion with the ideal answer. In third, fourth, tenth and eleventh rows there are comparisons of the RMS value. In the fifth to seventh and twelfth to fourteenth rows there are RMS values of the individual components of the inversions.}
\label{tab:GdG}
\centering
\begin{tabular}{p{0.1cm} p{3.68cm} c c c c}
\hline\hline
\multicolumn{6}{c}{GdG-type target function} \\
\multicolumn{6}{c}{cross-talk minimised} \\
\hline
 & & $v_x$ & $v_y$ & $v_z$ & $\delta c_s$ \\
\hline
1. & $\mathrm{corr}\left(\delta \hat{q}_{\alpha, \alpha}^{\mathrm{inv}} \mathrm{, id.\ ans.}\right)$ & 0.97 & 0.97 & 0.87 & 0.90 \\
2. & $\mathrm{corr}\left(\delta q_{\alpha}^{\mathrm{inv}} \mathrm{, id.\ ans.}\right)$ & 0.90 & 0.90 & 0.57 & 0.13 \\
\hdashline
3. & $\mathrm{RMS}\left(\delta \hat{q}_{\alpha, \alpha}^{\mathrm{inv}}\right)\,/\,\mathrm{RMS\left(id.\ ans.\right)}$ & 0.81 & 0.79 & 0.45 & 1.20 \\
4. & $\mathrm{RMS}\left(\delta q_{\alpha}^{\mathrm{inv}}\right)\,/\,\mathrm{RMS\left(id.\ ans.\right)}$ & 0.86 & 0.85 & 0.44 & 1.50 \\
\hdashline
5. & $\mathrm{RMS}\left(\delta \hat{q}_{\alpha,\alpha}^{\mathrm{inv}}\right)$ [\mps{}] & 55.11 & 41.63 & 21.61 & 31.18 \\
6. & $\mathrm{RMS}\left(\mathrm{cross\!-\!talk}\right)$ [\mps{}] & 0.51 & 0.52 & 0.73 & 3.82 \\
7. & $\mathrm{RMS}\left(\mathrm{noise}\right)$ [\mps{}] & 18.33 & 17.14 & 3.15 & 13.78 \\
\hline
\multicolumn{6}{c}{cross-talk ignored} \\
\hline
8. & $\mathrm{corr}\left(\delta \hat{q}_{\alpha,\alpha}^{\mathrm{inv}} \mathrm{, id.\ ans.}\right)$ & 0.99 & 0.99 & 0.94 & 0.91 \\
9. & $\mathrm{corr}\left(\delta q_{\alpha}^{\mathrm{inv}} \mathrm{, id.\ ans.}\right)$ & 0.93 & 0.94 & 0.36 & 0.42 \\
\hdashline
10. & $\mathrm{RMS}\left(\delta \hat{q}_{\alpha,\alpha}^{\mathrm{inv}}\right)\,/\,\mathrm{RMS\left(id.\ ans.\right)}$ & 0.92 & 0.93 & 0.47 & 1.25 \\
11. & $\mathrm{RMS}\left(\delta q_{\alpha}^{\mathrm{inv}}\right)\,/\,\mathrm{RMS\left(id.\ ans.\right)}$ & 0.96 & 0.96 & 0.49 & 1.88 \\
\hdashline
12. & $\mathrm{RMS}\left(\delta \hat{q}_{\alpha,\alpha}^{\mathrm{inv}}\right)$ [\mps{}] & 62.76 & 48.86 & 22.13 & 31.39 \\
13. & $\mathrm{RMS}\left(\mathrm{cross\!-\!talk}\right)$ [\mps{}] & 2.55 & 2.30 & 8.92 & 16.19 \\
14. & $\mathrm{RMS}\left(\mathrm{noise}\right)$ [\mps{}] & 18.01 & 15.98 & 5.25 & 13.17 \\
\hline
\end{tabular}
\end{table} 

In Section~\ref{subsect:p1} we performed an inversion for the sound-speed perturbations with a different target function and a different set of sensitivity kernels to improve the inversion. The improvements of such an approach are also apparent in Table~\ref{tab:Az_p1} (that should be compared to the last column in Table~\ref{tab:GdG}). The correlation coefficient of the inverted estimate is much larger than before, the random-noise level is smaller, and also the cross-talk pollution is smaller even in the case when it is not minimised.

\begin{table}
\caption{Correlation coefficients and RMS of inversions using the synthetic travel times. The Az target function was used. See Table \ref{tab:GdG} for details.}
\label{tab:Az_p1}
\centering
\begin{tabular}{p{0.1cm} l c}
\hline\hline
\multicolumn{3}{c}{Az-type target function} \\
\multicolumn{3}{c}{cross-talk minimised} \\
\hline
 & & $\delta c_s$ \\
\hline
1. & $\mathrm{corr}\left(\delta \hat{q}_{\alpha,\alpha}^{\mathrm{inv}} \mathrm{, id.\ ans.}\right)$ & 0.97 \\
1. & $\mathrm{corr}\left(\delta q_{\alpha}^{\mathrm{inv}} \mathrm{, id.\ ans.}\right)$ & 0.44 \\
\hdashline
3. & $\mathrm{RMS}\left(\delta \hat{q}_{\alpha,\alpha}^{\mathrm{inv}}\right)\,/\,\mathrm{RMS\left(id.\ ans.\right)}$ & 0.85 \\
4. & $\mathrm{RMS}\left(\delta q_{\alpha}^{\mathrm{inv}}\right)\,/\,\mathrm{RMS\left(id.\ ans.\right)}$ & 0.85 \\
\hdashline
5. & $\mathrm{RMS}\left(\delta \hat{q}_{\alpha,\alpha}^{\mathrm{inv}}\right)$ [\mps{}] & 48.49 \\
6. & $\mathrm{RMS}\left(\mathrm{cross\!-\!talk}\right)$ [\mps{}] & 1.27 \\
7. & $\mathrm{RMS}\left(\mathrm{noise}\right)$ [\mps{}] & 5.22 \\
\hline
\multicolumn{3}{c}{cross-talk ignored} \\
\hline
8. & $\mathrm{corr}\left(\delta \hat{q}_{\alpha,\alpha}^{\mathrm{inv}} \mathrm{, id.\ ans.}\right)$ & 0.98 \\
9. & $\mathrm{corr}\left(\delta q_{\alpha}^{\mathrm{inv}} \mathrm{, id.\ ans.}\right)$ & 0.48 \\
\hdashline
10. & $\mathrm{RMS}\left(\delta \hat{q}_{\alpha,\alpha}^{\mathrm{inv}}\right)\,/\,\mathrm{RMS\left(id.\ ans.\right)}$ & 0.81 \\
11. & $\mathrm{RMS}\left(\delta q_{\alpha}^{\mathrm{inv}}\right)\,/\,\mathrm{RMS\left(id.\ ans.\right)}$ & 0.71 \\
\hdashline
12. & $\mathrm{RMS}\left(\delta \hat{q}_{\alpha,\alpha}^{\mathrm{inv}}\right)$ [\mps{}] & 46.28 \\
13. & $\mathrm{RMS}\left(\mathrm{cross\!-\!talk}\right)$ [\mps{}] & 7.18 \\
14. & $\mathrm{RMS}\left(\mathrm{noise}\right)$ [\mps{}] & 4.77 \\
\hline
\end{tabular}
\end{table}

\begin{figure*}
        \sidecaption
        \includegraphics[width=10cm]{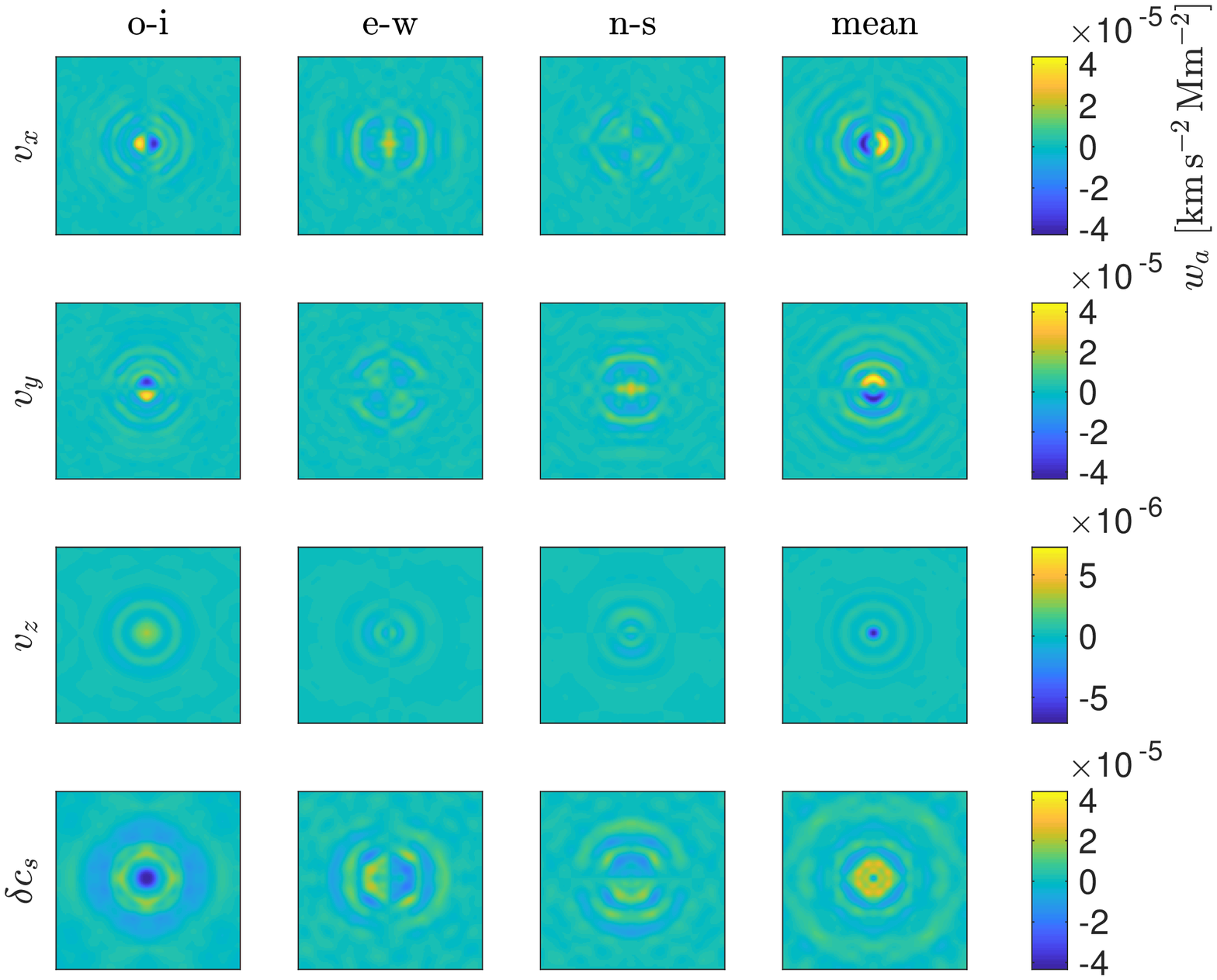}
        \caption{Examples of the weight functions of the inversion with the minimisation of the cross-talk and the GdG target function. Rows correspond to individual inversions. Columns correspond to individual averaging geometries. One can see the contributions of the mean geometry to the vector flows, especially to the vertical component.}
        \label{pic:vahy_GdG}
\end{figure*}

Moreover, in Fig. \ref{pic:vahy_GdG} we plot examples of the inversion weights for $\Delta \approx 19$ Mm. The weights are well localised around the central point.

\vspace{1cm}

\section{Conclusions}
We introduced the improved methodology of the SOLA time--distance inversions for local helioseismology. Our improvements consist of incorporating the mean travel-time geometry in the inversions and also of the use of the travel-time sensitivity kernels for all flow components and the sound-speed perturbations at once. This new methodology allows us to invert, for the first time, for the full value of the vertical velocity (not only for its variations as before) and to quantify the cross-talk between the vector flows and the sound-speed perturbations.

We validated our inversions by using the synthetic data based on the realistic numerical simulation of the solar convection provided by Matthias Rempel \citep{Rempel_2014,DeGrave_2014}. The use of the synthetic data allowed us to properly quantify all the inversion components and to gain empirical experience of the use of the improved pipeline. To summarise the results:

\begin{itemize}
\item For $v_x$ and $v_y$ (the horizontal flow) it is not necessary to minimise the cross-talk (at least near the surface). Our methodology essentially does not constitute an improvement to what has been used recently by other authors. We only point out that the targeted RMS of the random noise should be 20~\mps{} or less if one inverts the near-surface flows using travel times averaged over a day or so. 

\item For $v_z$ (the vertical flow) it is essential to minimise the cross-talk, otherwise the results are nullified by the leakage mainly from the horizontal components. The leakage from the sound-speed perturbations is acceptably small. Our findings confirm the results seen in the past. The RMS of the random noise should be lower than 2~\mps{} for the near-surface flows (again, for one-day averaged travel times). An obvious advantage of our new methodology is that one can also quantify the large-scale (with scales larger than the typical horizontal extent of the averaging kernel) component of the vertical flow and not only its horizontal variations as before. The inversion is sensitive also to the large-scale offset because of the incorporation of the mean travel-time kernels. Also, it seems that the incorporation of the mean travel times helps to regularise the cross-talk contribution. By using the improved methodology, we achieved significant improvements in inversions for the vertical flow.

\item At first sight it seems that the minimisation of the cross-talk in the case of the inversion for $\delta c_s$ does not improve the results. This is because the leakage component from the other quantities is positively correlated with the sound-speed perturbations and thus the leakage ``helps'' to ``measure'' $\delta c_s$. Such an approach leads to an overestimation of the magnitude of the sound-speed variations by almost a factor of two in the near-surface layers. Such an inversion, however, is not the inversion for $\delta c_s$. When the cross-talk minimisation is introduced, the agreement with the ideal answer decreases, which is in part due to the worse fit of the target function by the averaging kernel, and also by the lower signal-to-noise ratio achieved for the travel times averaged over 24 hours. The case we present here cannot be further improved because only the $f$ mode, which is not particularly sensitive to the sound-speed perturbations, was used. The reliability improves by a large extent when the first acoustic ridge is used in a similar setup. 
\end{itemize}

\begin{acknowledgements}
D.K. is supported by the Grant Agency of Charles University under grant No. 532217. M.\v{S}. is supported by the project RVO:67985815 and the grant project 18-06319S awarded by Czech Science Foundation. We are grateful to Matthias Rempel for providing us with the snapshot of the numerical simulation. The sensitivity kernels were computed by the code {\sc Kc3} kindly provided by Aaron Birch. The authors would like to thank the anonymous referee for the valuable comments and suggestions, which greatly improved the quality of the paper. This research has made use of NASA Astrophysics Data System Bibliographic Services. Author contributions: M.\v{S}. and D.K. designed the
research. D.K. implemented and performed the research within his Ph.D projects under the supervision of M.\v{S}. D.K. drafted the paper and both authors contributed to the final manuscript. 
\end{acknowledgements}

\bibliographystyle{aa} 
\bibliography{BIBL}

\end{document}